\pdfoutput=1
\documentclass[preprint,11pt]{elsarticle}
\usepackage{graphicx}
\usepackage{verbatim}
\usepackage{epsfig}
\usepackage{titlesec}
\usepackage{amsmath, amsfonts, amssymb}
\usepackage{mathrsfs}
\usepackage{xcolor}
\usepackage{rotating}
\usepackage{mathtools}
\usepackage{multirow}
\usepackage{amsthm}
\usepackage{esvect}
\usepackage{multicol}
\usepackage{subcaption}
\usepackage{algorithm}
\usepackage{algpseudocode}
\usepackage{geometry,longtable,booktabs, caption}
\usepackage{comment}
\usepackage{float}
\usepackage{placeins}
\usepackage{enumerate}
\usepackage{bm}
\usepackage{natbib}
\usepackage[hidelinks]{hyperref}
\usepackage{comment}
\setcitestyle{authoryear,open={(},close={)}} %引用设置

\titleformat{\subsection}
  {\normalfont\bfseries}
  {\thesubsection}
  {1em}
  {}
\titleformat{\subsubsection}
  {\normalfont\normalsize\bfseries}
  {\thesubsubsection}
  {1em}
  {}
\def\be{\begin{equation}}
\def\ee{\end{equation}}
\def\bq{\begin{eqnarray}}
\def\eq{\end{eqnarray}}

\makeatletter
\def\ps@pprintTitle{%
  \let\@oddhead\@empty
  \let\@evenhead\@empty
  \def\@oddfoot{\reset@font\hfil\thepage\hfil}
  \let\@evenfoot\@oddfoot
}
\makeatother

\begin{document}
\begin{frontmatter}
	
\title{Weekend Crypto Volatility and Monday Market Risk}

\author{Pedram Ebadollahyvahed\footnotemark[1], Nneka Umeorah\footnotemark[1]}
\address{\footnotemark[1]School of Mathematics, Cardiff University, Cardiff CF24 4AG, United Kingdom\\
Emails: EbadollahyvahedP@cardiff.ac.uk; umeorahn@cardiff.ac.uk\\
\textbf{Corresponding author}: Pedram Ebadollahyvahed}

\begin{abstract}
\noindent Cryptocurrencies trade continuously while major traditional markets close over the weekend. We examine whether weekend cryptocurrency volatility predicts Monday variance in equity, commodity and fixed-income markets. Using daily range-based variance measures for Bitcoin, Ether, XRP, the S\&P 500, gold and a US Treasury ETF over 2017–2026, we augment heterogeneous autoregressive models with a common weekend cryptocurrency factor. Predictive content is weak before 2024 but significant thereafter. In the post-2024 sample, a one-standard-deviation increase in the factor is associated with 27–54\% higher fitted Monday variance. Adding the factor reduces out-of-sample QLIKE losses by 11–22\%, although statistical evidence varies across markets. Friday cryptocurrency volatility does not reproduce the result, and coin-specific variation adds no significant predictive information beyond the common factor. An endogenous procedure estimates a break in late 2023, during growing anticipation of US spot-Bitcoin ETF approval. However, ETF-flow interactions provide no evidence of amplification through greater measured Monday flow activity, leaving the mechanism unresolved. The findings suggest that calendar-specific volatility measures reveal predictive information that aggregate connectedness measures may obscure.
\medskip

\end{abstract}
\end{frontmatter}

\section{Introduction}\label{sec:intro}
The trading calendars of cryptocurrency and traditional markets do not coincide. Equities, commodities and bonds close over the weekend, while cryptocurrencies trade continuously. Information arriving during this period can therefore be incorporated into cryptocurrency prices before traditional markets reopen. Most empirical studies address this calendar mismatch by aligning cryptocurrency and traditional-market observations to a common trading calendar. We instead exploit the mismatch as a source of temporal identification: cryptocurrency markets can respond to weekend information while traditional markets remain closed, allowing us to test whether volatility accumulated in cryptocurrency during this interval contains predictive information for traditional-market volatility when trading resumes.

Using this timing asymmetry, we find that weekend cryptocurrency volatility contains predictive information for volatility in traditional markets on Monday. Specifically, higher weekend cryptocurrency variance predicts higher Monday variance in the S\&P 500, gold and a US Treasury bond ETF, even after controlling for each asset’s recent volatility and for cryptocurrency variance on the preceding Friday. Importantly, this relationship is not evident earlier in the sample. It emerges around 2023–24 and is statistically significant across all nine cryptocurrency–traditional asset pairs considered in this study during the post-2024 period.

The timing coincides with an important period of institutional change in cryptocurrency markets. On 10~January~2024, the US Securities and Exchange Commission approved rule changes permitting the listing and trading of eleven
spot-Bitcoin exchange-traded products in the United States, following repeated rejections and a court ruling vacating its earlier denial
of Grayscale's proposal \citep{sec2024}. These products expanded access to Bitcoin exposure through regulated US securities exchanges for institutional and retail investors. A growing literature documents increased integration between cryptocurrency and traditional markets around this period \citep{babalos2025,bukaita2026,hong2025}. Whereas this literature primarily characterises changes in aggregate integration, we examine whether a specific calendar-based predictive relationship emerged around the same period. An endogenous structural-break procedure places the change in late 2023, broadly coinciding with the period in which spot-Bitcoin ETF approval became increasingly anticipated.

Several findings support the interpretation that the predictive relationship is specifically associated with weekend cryptocurrency volatility rather than cryptocurrency volatility more generally. First, weekend variance remains statistically significant when Friday cryptocurrency variance is included as a control, whereas Friday variance alone has little predictive content for Monday volatility. Second, the relationship is driven primarily by a common cryptocurrency-market component. When each coin’s weekend variance is decomposed into common and coin-specific components, the common factor significantly predicts Monday variance across all three traditional assets, while the coin-specific components do not. Third, the relationship is economically and forecast relevant. In the post-2024 sample, a one-standard-deviation increase in the weekend cryptocurrency factor is associated with approximately 28\% higher Monday variance for the S\&P 500, 54\% for gold and 27\% for the Treasury ETF, while adding the factor to the HAR benchmark reduces out-of-sample QLIKE loss by 11–22\%. Fourth, the main finding is robust to several alternative specifications, including controlling for the pre-weekend VIX, estimating a date-clustered panel specification, excluding the most influential observation, and allowing the timing of the structural change to be determined endogenously.

The calendar-specific nature of the relationship also helps explain why it is not apparent in conventional measures of market connectedness. Applying the Diebold–Yilmaz connectedness framework to the same six assets produces relatively weak volatility transmission from individual cryptocurrencies to the traditional assets, while the rolling connectedness index shows no comparable change around the 2023–24 period. This contrast suggests that transmission concentrated around a particular part of the trading calendar may be diluted when connectedness is measured using daily observations pooled across all weekdays. More generally, the result highlights the value of explicitly accounting for differences in trading calendars when studying information flows between continuously traded and periodically closed markets.

Finally, we also examine whether spot cryptocurrency ETF flows provide a potential mechanism linking weekend cryptocurrency volatility to Monday volatility in traditional markets. If ETF activity were the primary conduit, we would expect the relationship to strengthen when fund flows are larger. We find no evidence of this pattern: the interaction between weekend cryptocurrency variance and ETF flows is statistically insignificant for the S\&P 500 and gold and negative for the Treasury ETF. Thus, although the relationship emerges during a period of increasing cryptocurrency institutionalisation, our results do not support ETF flows as the mechanism driving it. The underlying mechanism therefore remains unresolved.

The study most closely related to ours is \citet{mourey2025}, who show that weekend cryptocurrency returns predict Monday equity returns and date a structural change in that relationship to the 2022 LUNA collapse. Our analysis differs along three dimensions. First, we examine volatility rather than returns, shifting the focus from return predictability to measuring and forecasting market risk. Second, we consider three traditional asset classes: equities, commodities and fixed income, rather than equities alone. Third, we examine the timing and potential mechanisms underlying the emergence of the volatility relationship, including structural-break analysis and ETF-flow interactions. The distinction between returns and volatility is economically important because volatility forecasts enter risk management, margin requirements and value-at-risk calculations. Our analysis also relates to the classical evidence of \citet{frenchroll1986} that return variance differs substantially between trading and non-trading periods. Continuous cryptocurrency trading provides a setting in which information can be reflected in one market while traditional markets remain closed.

This paper makes four contributions.
\begin{itemize}
\item \textbf{A new empirical finding.} We document a calendar-specific relationship in which weekend cryptocurrency volatility predicts Monday volatility across equities, commodities and fixed income, emerging primarily around 2023–24.
\item \textbf{A methodological contribution.} We show that accounting explicitly for differences in trading calendars can reveal predictive relationships that may be less apparent in aggregate connectedness measures.
\item \textbf{A forecasting contribution.} We compare conventional and machine-learning volatility models and find that tuned machine-learning methods do not consistently outperform the HAR benchmark. However, adding the weekend cryptocurrency factor improves HAR forecasts.
\item \textbf{Evidence on the mechanism.} ETF-flow interactions provide no evidence that spot-ETF flows explain the weekend relationship, suggesting that its emergence cannot be attributed to this specific channel.
\end{itemize}

The remainder of the paper is organised as follows. Section~\ref{sec:lit} reviews the relevant literature.
Section~\ref{sec:method} presents the empirical methodology, including the volatility measures, forecasting framework and the weekend-channel specification. Section~\ref{sec:data} describes the data and the choice of assets. Section~\ref{sec:results} examines the emergence of the weekend
relationship, distinguishes weekend from Friday and coin-specific information, and assesses its economic magnitude and forecasting value. It also presents
robustness checks, ETF-flow interactions, and complementary forecasting and connectedness analyses. Section~\ref{sec:conc} concludes the study.

\section{Related literature}\label{sec:lit}
This study builds on research on volatility measurement and forecasting, cryptocurrency volatility, cross-market connectedness, and calendar effects in financial markets. Together, these areas provide the theoretical and empirical background for examining whether cryptocurrency volatility observed while traditional markets are closed contains predictive information for volatility when those markets reopen.

\subsection{Measuring and forecasting volatility}
Volatility is latent and must therefore be estimated from observed prices. Financial returns are characterised by heavy tails and volatility clustering \citep{mandelbrot1963,fama1965,cont2001}, features that motivated the development of ARCH and GARCH models \citep{engle1982,bollerslev1986} and their asymmetric extensions \citep{nelson1991,gjr1993}. With the availability of high-frequency data, realised-volatility measures constructed from intraday returns provided more informative ex-post measures of price variation than squared daily returns \citep{andersenbollerslev1998,abde2001,abdl2003,bns2002}.

The heterogeneous autoregressive (HAR) model of \citet{corsi2009} provides a parsimonious framework for forecasting realised volatility using its daily, weekly and monthly components. Its ability to capture volatility persistence while remaining relatively simple has made it a widely used benchmark for volatility forecasting. When intraday data are unavailable, range-based estimators provide an alternative by extracting information from daily open, high, low and close (OHLC) prices \citep{parkinson1980,garmanklass1980,rogerssatchell1991}. These estimators are particularly relevant to our analysis because our volatility measures are constructed from daily OHLC data.

Forecast evaluation also requires care because observed volatility measures are proxies for the underlying latent variance. Loss functions such as QLIKE are robust to certain forms of volatility-proxy noise \citep{patton2011}, while the Diebold-Mariano test provides a formal comparison of predictive accuracy across competing forecasts \citep{dm1995}. Previous evidence also shows that greater model complexity does not necessarily translate into superior volatility forecasts \citep{hansenlunde2005,dudek2024}. We therefore use the HAR model as a parsimonious benchmark and assess whether incorporating weekend cryptocurrency volatility provides additional predictive information for Monday volatility.

\subsection{Cryptocurrency volatility and traditional financial assets}

Cryptocurrencies have developed from an alternative digital payment system into a widely
traded class of financial assets. Their market behaviour nevertheless differs in important
respects from that of traditional assets. \citet{corbet2019} review the development of
cryptocurrency markets and document their historically low correlations with conventional
financial assets, a feature that initially motivated interest in their potential diversification
benefits. A particularly distinctive characteristic is their high and persistent volatility.
For Bitcoin, \citet{katsiampa2017} shows that GARCH-type models capture important features
of the volatility dynamics, while \citet{baurdimpfl2018} document an asymmetric response
to shocks that differs from the pattern commonly observed in equity markets: positive shocks
can generate stronger volatility responses than negative shocks.

Research has also examined the broader financial role of cryptocurrencies. Questions of informational efficiency have received considerable attention \citep{urquhart2016}, alongside
debate over whether cryptocurrencies behave primarily as speculative assets, alternative means of exchange, diversifiers or hedges \citep{dyhrberg2016,bouri2017}. Gold is particularly relevant in this context because of its established role as a store of value and safe-haven asset \citep{baurlucey2010}, and comparisons between Bitcoin and gold have therefore become
common in the cryptocurrency literature. These studies show that the relationship between cryptocurrencies and traditional assets cannot be characterised solely by their individual return or volatility dynamics; their degree of co-movement and potential for cross-market information transmission are also important.

Another feature distinguishes cryptocurrencies from the traditional assets considered in this paper: cryptocurrency markets operate continuously, including when major traditional markets are closed. Although trading is available around the clock, activity is not uniform across time; \citet{Jahanshahloo2022} document pronounced time-of-day effects in Bitcoin trading and network activity, strongly influenced by conventional US trading hours, with these effects diminishing at weekends. The absence of scheduled weekend closure nevertheless creates periods when cryptocurrency prices can adjust to new information while major equity, bond, and commodity markets are not trading. This difference in trading calendars has become increasingly relevant as cryptocurrency markets have developed stronger links with the wider financial system. Using return and volatility connectedness measures, \citet{iyer2023} document increasing spillovers between crypto assets and traditional financial markets over time, particularly during periods of market turbulence. More directly related to the timing considered here, \citet{mourey2025} exploit the difference between continuous cryptocurrency trading and weekend equity-market closure and show that weekend cryptocurrency returns contain predictive information for Monday equity returns, with the relationship strengthening following the LUNA collapse in 2022. Taken together, these findings motivate examining whether the same calendar-based relationship extends to volatility and, beyond equities, to commodity and fixed-income markets.

\subsection{Volatility connectedness and post-2024 integration}

Cross-market volatility transmission is commonly studied using measures of spillovers and connectedness. A prominent approach is the Diebold-Yilmaz framework \citep{dy2012,dy2014}, which uses generalised forecast-error variance decompositions to quantify the proportion of variation in one market associated with shocks originating in others. The framework has subsequently been extended to examine connectedness across different frequencies \citep{bk2018} and to allow relationships between markets to evolve
over time \citep{acg2020}. Applications to cryptocurrency markets provide evidence that their links with traditional financial assets have strengthened over time and that the magnitude and direction of volatility spillovers can vary during periods of market stress \citep{imf2022,iyer2023}. More recent studies examine how these relationships have evolved alongside the increasing institutionalisation of cryptocurrency markets. Around the
introduction of US spot-Bitcoin exchange-traded products in January 2024, evidence points to changes in Bitcoin's co-movement with equity markets and in its volatility dynamics
\citep{babalos2025,hong2025}, while \citet{bukaita2026} document changes in the spillover structure linking Bitcoin, Ether, gold and US equities.

These studies provide evidence of an evolving relationship between cryptocurrency and traditional financial markets, but connectedness is generally measured using observations
aggregated over time and is designed to characterise the overall magnitude and direction of cross-market spillovers \citep{dy2012,dy2014,bk2018}. This may leave calendar-specific
relationships less visible when predictive information concentrates in particular trading periods. The distinction is especially relevant for cryptocurrency markets, where
trading continues outside conventional market hours and through weekends, despite pronounced time-of-day patterns in trading and network activity \citep{Jahanshahloo2022}. Indeed, \citet{mourey2025} show that exploiting the weekend closure of equity markets reveals predictive information in weekend cryptocurrency returns for Monday equity returns. These
findings motivate a complementary approach to aggregate connectedness analysis: we examine whether cryptocurrency volatility accumulated specifically over the weekend contains
predictive information for traditional-market volatility on Monday, and subsequently compare this calendar-specific relationship with conventional connectedness measures in
Section~\ref{sec:conn}.

\subsection{Calendar effects and information transmission}

Calendar effects have long been documented in financial markets. Early studies identify systematic differences in equity returns across trading days, including the weekend effect
\citep{cross1973,french1980}. \citet{frenchroll1986} extend this perspective to volatility, showing that return variance differs substantially between trading and non-trading periods and highlighting the importance of market opening and closing for incorporating information into prices. Calendar regularities are also present in cryptocurrency markets despite their continuous trading structure. \citet{caporale2019}, for example, document day-of-the-week effects in cryptocurrency returns, suggesting that continuous trading does not eliminate systematic temporal patterns. More directly related to our analysis,
\citet{mourey2025} exploit the difference between continuous cryptocurrency trading and weekend equity-market closure and show that weekend cryptocurrency returns predict Monday
equity returns, with a structural change in the relationship around the 2022 LUNA collapse. Their analysis establishes the relevance of the weekend trading-calendar mismatch for return
predictability. We examine whether a corresponding relationship exists in volatility: whether cryptocurrency variance accumulated over the weekend contains predictive information for Monday volatility in traditional financial markets. In doing so, we shift the focus from return predictability to market-risk forecasting, extend the analysis beyond equities to commodity and fixed-income markets, and examine when the volatility relationship emerges
and the mechanisms that may account for its evolution.

\section{Methodology}\label{sec:method}
This section presents the methodology used to examine whether cryptocurrency volatility over the weekend provides additional information about Monday variance in traditional financial markets. We first construct daily variance proxies from OHLC prices and specify a heterogeneous autoregressive (HAR) model as the forecasting benchmark. An expanding-window forecasting framework is then used to assess predictive performance using only information available at each forecast date. The central analysis augments the HAR benchmark with weekend and Friday cryptocurrency variance factors to distinguish the predictive contribution of weekend activity from that of preceding Friday activity. Further analyses examine changes in this relationship over time, distinguish common cryptocurrency-market variation from coin-specific variation, and assess whether ETF flow activity is associated with the strength of the relationship. The section concludes with the forecast evaluation criteria and statistical inference procedures.

\subsection{Measuring daily variance}\label{sec:rvmeasure}

Volatility is not directly observed and must be estimated from price data. We construct daily variance proxies using each asset's opening, highest, lowest and closing prices, denoted by $O_{i,t}$, $H_{i,t}$, $L_{i,t}$ and $C_{i,t}$, respectively. Here, $i$ indexes assets and $t$ denotes the date of the daily observation. Range-based estimators use information about price variation within the observation window that is not captured by squared close-to-close returns.
Our primary measure is the \citet{garmanklass1980} estimator:
\begin{equation}
\widehat{\sigma}_{i,t}^{2,\mathrm{GK}}=\frac{1}{2}
\left[\ln\left(\frac{H_{i,t}}{L_{i,t}}\right)\right]^2-(2\ln 2-1)
\left[\ln\left(\frac{C_{i,t}}{O_{i,t}}\right)\right]^2.
\label{eq:gk}
\end{equation}
As robustness checks, reported in \ref{appendix}, we also compute the \citet{parkinson1980} estimator,
\begin{equation}
\widehat{\sigma}_{i,t}^{2,\mathrm{P}}=\frac{1}{4\ln 2}
\left[\ln\left(\frac{H_{i,t}}{L_{i,t}}\right)\right]^2,
\label{eq:park}
\end{equation}
and the \citet{rogerssatchell1991} estimator,
\begin{equation}
\begin{aligned}
\widehat{\sigma}_{i,t}^{2,\mathrm{RS}}
={}&
\ln\left(\frac{H_{i,t}}{C_{i,t}}\right)
\ln\left(\frac{H_{i,t}}{O_{i,t}}\right)+
\ln\left(\frac{L_{i,t}}{C_{i,t}}\right)
\ln\left(\frac{L_{i,t}}{O_{i,t}}\right).
\end{aligned}
\label{eq:rs}
\end{equation}
The Parkinson estimator uses the high-low range alone, whereas the
Rogers-Satchell estimator accommodates non-zero drift under its
underlying diffusion assumptions.

Before constructing the analysis series, we clean the daily
Garman-Klass volatility estimates. Zero estimates are treated
as missing. Separately for each asset, non-missing volatility
estimates below their 0.5th percentile are replaced by that
threshold. Missing observations are then forward-filled, with
backward-filling used only for any remaining leading gaps.
The cleaned volatility estimates are squared to obtain the
daily variance proxy, denoted by $RV_{i,t}$. Thus, unless
otherwise stated, $RV_{i,t}$ refers to the cleaned
Garman-Klass variance proxy.

We work with its logarithmic transformation:
\begin{equation}
v_{i,t}=\ln(RV_{i,t}+\epsilon),
\qquad \epsilon=10^{-12}.
\label{eq:log_variance}
\end{equation}
The positive constant $\epsilon$ provides a numerical safeguard
for the logarithmic transformation. Working in logarithms
reduces right-skewness and compresses large variance observations. These estimators measure price variation within the observation window represented by each daily bar. For cryptocurrencies, this window covers a calendar day; for traditional assets, it follows the trading session represented by the data source. The estimators do not directly include the return between the previous session's close and the current session's open. Consequently, Monday $RV_{i,t}$ measures within-session variation rather than the total price adjustment between Friday's close and Monday's close. Also, where annualised volatility is reported, we use
$\sqrt{252\,RV_{i,t}}$ for traditional assets and
$\sqrt{365\,RV_{i,t}}$ for cryptocurrencies, reflecting their
different trading calendars.

\subsection{The HAR benchmark}\label{sec:har}

We use the HAR model of \citet{corsi2009} as the benchmark for forecasting daily log variance. The model captures persistence over daily, weekly and monthly horizons through a parsimonious combination of lagged log variance and its historical averages.

Let $v_{i,t}=\ln(RV_{i,t}+\epsilon)$, as defined in
Section~\ref{sec:rvmeasure}, and define
\begin{equation}
\bar v_{i,t-1:t-h}
=
\frac{1}{h}\sum_{j=1}^{h}v_{i,t-j}.
\label{eq:har_average}
\end{equation}
Here, $t-j$ denotes the $j$th available daily observation preceding
the observation indexed by $t$. For traditional assets, these lags
refer to trading sessions; for cryptocurrencies, they refer to
calendar days. Thus, for a traditional asset observed on Monday,
$v_{i,t-1}$ normally denotes the preceding Friday's log variance.

The HAR specification is
\begin{equation}
v_{i,t}=\beta_0+\beta_d v_{i,t-1}+\beta_w \bar v_{i,t-1:t-5}
+\beta_m \bar v_{i,t-1:t-22}+\varepsilon_{i,t},
\label{eq:har}
\end{equation}
where $\beta_0$ is the intercept, $\beta_d$, $\beta_w$ and $\beta_m$ measure dependence on the daily, weekly and monthly components, respectively, and $\varepsilon_{i,t}$ is the regression disturbance. The model is estimated separately for each asset by ordinary least squares; the asset index on the coefficients is suppressed for notational simplicity.

The five- and twenty-two-observation averages follow the conventional HAR specification. They approximate a trading week and a trading month for traditional assets. For cryptocurrencies, which trade continuously, they represent five- and twenty-two-calendar-day windows. Retaining these observation counts provides a common benchmark specification across assets, although the corresponding calendar horizons differ.

The HAR model serves two purposes. First, it provides a benchmark
against which the out-of-sample performance of alternative forecasting models is assessed. Second, its daily, weekly and monthly components control for the traditional asset's own variance history in the weekend-channel regressions. The weekend cryptocurrency factor is therefore assessed for the additional predictive information it provides beyond these persistence controls.

\subsection{Expanding-window forecasting design}
\label{sec:walkforward}
We assess forecasting performance using an expanding-window
walk-forward design. For each forecast date $t$, a one-step-ahead
forecast of $v_{i,t}$ is generated using only information available at the forecast origin. Lagged predictors are constructed as defined in equation~\ref{eq:har}.

For the general forecasting comparison, HAR and the tree-based models are re-estimated every $21$ observations, whereas the LSTM is re-estimated every $63$ observations. Between re-estimation dates, parameters
remain fixed while predictors are updated using newly available
observations. For the Monday comparison between HAR and
HAR+$W_t$, both models are instead re-estimated before each
Monday forecast using all preceding Monday observations in
the estimation window. The HAR predictors are constructed
from the asset's daily trading history, and $W_t$ uses the
completed Saturday and Sunday cryptocurrency observations.

In both exercises, the estimation window expands over time.
Any feature scaling and hyperparameter selection use only
the corresponding training data.

\subsection{The weekend volatility channel}\label{sec:weekend}

We examine whether cryptocurrency variance observed over the
weekend provides additional predictive information about Monday
variance in traditional financial markets. The analysis augments
the HAR benchmark with weekend and Friday cryptocurrency variance
factors. The HAR terms control for the traditional asset's own
variance history, while the Friday factor helps distinguish the
predictive contribution of weekend activity from that of preceding
Friday activity. We refer to this predictive relationship as the
weekend volatility channel.

\subsubsection{Constructing the weekend and Friday factors}

We construct a weekend variance proxy by summing the Saturday
and Sunday daily variance estimates. For cryptocurrency $c$ and
a Monday dated $t$, define
\begin{equation}
RV^{\mathrm{wk}}_{c,t}=RV_{c,t-1}+RV_{c,t-2},
\qquad
RV^{\mathrm{fri}}_{c,t}=RV_{c,t-3},
\label{eq:wkrv}
\end{equation}
where, in this construction, the offsets denote calendar days:
$t-1$ is Sunday, $t-2$ is Saturday and $t-3$ is Friday. This
calendar-day convention applies to the cryptocurrency factors;
the HAR lags for traditional assets continue to refer to preceding
available trading observations, as defined in Section~\ref{sec:har}. The sum $RV^{\mathrm{wk}}_{c,t}$ approximates accumulated weekend variance using the range-based daily proxies defined in Section~\ref{sec:rvmeasure}.

To capture variation shared across the three cryptocurrencies,
we construct equally weighted market-level factors by averaging
their logged variance proxies:
\begin{equation}
W_t
=
\frac{1}{N_c}\sum_{c=1}^{N_c}
\ln\!\left(RV^{\mathrm{wk}}_{c,t}+\epsilon\right),
\qquad
F_t
=
\frac{1}{N_c}\sum_{c=1}^{N_c}
\ln\!\left(RV^{\mathrm{fri}}_{c,t}+\epsilon\right),
\label{eq:wkfactor}
\end{equation}
where $N_c=3$ and $\epsilon=10^{-12}$, consistent with
Section~\ref{sec:rvmeasure}. Thus, $W_t$ measures common weekend cryptocurrency
variance in logarithmic units, while $F_t$ is its Friday
counterpart. The daily variance estimates are summed before
taking logarithms to construct the weekend factor.

\begin{figure}[H]
\centering
\includegraphics[width=\textwidth]{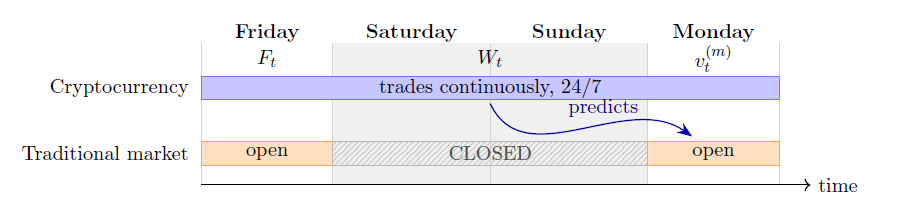}
\caption{Timing of the weekend predictive test. The factor $W_t$
uses Saturday and Sunday cryptocurrency variance observed during
the regular US equity-market closure to predict Monday variance.
Related instruments may trade before Monday's cash-market opening.}
\label{fig:ident}
\end{figure}

The distinction between the two factors is what identifies the channel, and Figure~\ref{fig:ident} sets out the
timing on which it rests. Cryptocurrency daily bars close at 00:00~UTC, so the Saturday and Sunday bars that enter
$RV^{\mathrm{wk}}_{c,t}$ together span Saturday 00:00 to Monday 00:00~UTC. The regular US equity session closes on Friday at 21:00 UTC during US standard time or 20:00 UTC during US daylight saving time, and reopens on Monday at 14:30 or 13:30 UTC, respectively, so the whole of the weekend window lies inside the period
during which the traditional market was closed: nothing that enters $W_t$ could have been traded before Monday's
open. The Friday bar, by contrast, spans a calendar day on which the traditional market was open, so the
information it carries could be priced on Friday itself.\footnote{The final three hours of the Friday bar during US standard time (21:00–00:00 UTC), or four hours during US daylight saving time (20:00–00:00 UTC), fall after the regular US equity session has closed. This slice makes $F_t$ slightly
weekend-like and therefore works against the distinction we draw; a finding that $W_t$ predicts while $F_t$ does
not is conservative on this account.} If the traditional market reacts on Monday to information it could not
trade at the weekend, $W_t$ should predict Monday variance; if the effect merely reflects recent cryptocurrency
turbulence in general, $F_t$ should predict it just as well. Gold is a partial exception, since COMEX futures
reopen on Sunday evening and the last hours of the Sunday bar are therefore tradable; we return to this in
Section~\ref{sec:reopen}.

\subsubsection{The HAR-augmented predictive regression}

For traditional asset $m$ and Monday $t$, we augment the HAR
benchmark in \eqref{eq:har} with the weekend and Friday
cryptocurrency variance factors:
\begin{equation}
v^{(m)}_t=\alpha+\beta_d v^{(m)}_{t-1}+\beta_w \bar v^{(m)}_{t-1:t-5}+\beta_{mo} \bar v^{(m)}_{t-1:t-22}+\gamma W_t+\delta F_t+\varepsilon_t.
\label{eq:weekend}
\end{equation}
Here, $v^{(m)}_t=v_{m,t}$ denotes the log variance of traditional
asset $m$, using the transformation defined in Section~\ref{sec:rvmeasure}.
The regression is estimated separately for each traditional
asset using Monday observations; asset indices on the
coefficients and disturbance are suppressed for simplicity.
The HAR lags refer to preceding available trading observations,
so $v^{(m)}_{t-1}$ normally denotes the preceding Friday's
log variance. The five- and twenty-two-observation averages
are constructed from the asset's daily trading history,
rather than from Monday observations alone.

The coefficient of interest is $\gamma$, which measures the
additional predictive association between weekend cryptocurrency
variance and Monday log variance after controlling for the
asset's own variance history and Friday cryptocurrency variance.
The null hypothesis of no additional weekend predictive
content is $H_0:\gamma=0$.

Because $W_t$ and $F_t$ are correlated, their joint inclusion
may increase the uncertainty of the coefficient estimates.
We therefore estimate \eqref{eq:weekend} in three forms:
the joint model as written, a weekend-only model omitting
$F_t$, and a Friday-only model omitting $W_t$. The separate
models assess each factor conditional on the HAR controls,
while the joint model assesses each factor conditional on
both the HAR controls and the other cryptocurrency factor.

We also assess whether the weekend coefficient exceeds the
Friday coefficient through the one-sided contrast
\begin{equation}
H_0:\gamma\leq\delta
\qquad \text{against} \qquad
H_1:\gamma>\delta.
\label{eq:wald}
\end{equation}
This test is computed from the joint model using the robust
covariance estimator described in Section~\ref{sec:evaluation}, with the null evaluated at the boundary $\gamma=\delta$.

Evidence for the weekend volatility channel is assessed
through these coefficient estimates and tests, together with
the out-of-sample forecasting results. Significance in the
weekend-only model does not by itself establish predictive
content beyond Friday activity, and the regressions do not
establish causal transmission.

\subsubsection{Dating the emergence of the channel}

To examine when the channel emerged, we estimate $\gamma$ within
disjoint sub-periods and rolling windows, and interact $W_t$ with
indicators for candidate break dates. We also estimate the break
date rather than impose it in advance. For each candidate date
$\tau$ in the trimmed sample interval $[\tau_1,\tau_2]$, we compute the Wald statistic $\mathcal{W}(\tau)$ for a change in $\gamma$ and define
\begin{equation}
\hat{\tau}=\arg\max_{\tau\in[\tau_1,\tau_2]}\mathcal{W}(\tau),
\qquad \sup\text{-}\mathcal{W}=\max_{\tau\in[\tau_1,\tau_2]}\mathcal{W}(\tau).
\label{eq:supwald}
\end{equation}
We exclude the first and last $15\%$ of observations from the
candidate break dates and evaluate the resulting statistic
against the critical values of \citet{andrews1993}, which
account for the break date being estimated rather than known.

\subsubsection{Separating market-level from coin-specific transmission}

To examine whether the channel reflects common cryptocurrency-market
variation or coin-specific variation, we decompose each coin's
logged weekend variance into a component explained by $W_t$ and
a residual:
\begin{equation}
\ln\!\left(RV^{\mathrm{wk}}_{c,t}+\epsilon\right)
=
a_c+b_cW_t+u_{c,t}.
\label{eq:orth}
\end{equation}
The coefficient $b_c$ measures the coin's loading on the common
factor, while $u_{c,t}$ captures variation in its logged weekend
variance not explained by $W_t$. We estimate this decomposition
by ordinary least squares.

We then augment \eqref{eq:weekend} with each estimated residual
$\hat u_{c,t}$ separately, retaining $W_t$ and the other controls.
Because $W_t$ is the average of the same logged weekend variances,
the residuals sum to zero when estimated over a common sample
and cannot all be included jointly.

If $W_t$ remains significant while the coin-specific residuals
provide no additional predictive information, the results are
consistent with a predominantly market-level channel. A significant
residual coefficient instead indicates additional predictive
information associated with that coin beyond the common factor.

\subsubsection{Testing the exchange-traded-fund flow mechanism}

A candidate mechanism is that institutional ETF activity is associated with the transmission of weekend cryptocurrency information into Monday trading.
The variable $\mathrm{Flow}_t$ is constructed by summing
the absolute aggregate daily net flows into US spot
Bitcoin and Ether ETFs, then subtracting the sample
mean and dividing by the sample standard deviation.
This measure captures the magnitude of aggregate net
flows rather than ETF trading volume.

The interaction between $W_t$ and $\mathrm{Flow}_t$
tests whether the weekend relationship is stronger
on Mondays with greater measured flow activity:
\begin{equation}
v^{(m)}_t=\alpha+\sum_j\beta_j x_{j,t}+\gamma W_t+\delta F_t +\phi\,\mathrm{Flow}_t +\psi\left(W_t\times\mathrm{Flow}_t\right)
+\varepsilon_t.
\label{eq:flow}
\end{equation}
Here, $\sum_j\beta_j x_{j,t}$ denotes the daily, weekly and monthly HAR terms in \eqref{eq:weekend}. The coefficient $\phi$ captures the association between ETF flow activity and Monday log variance, while $\psi$ measures how the weekend coefficient varies with flow activity. The marginal association with $W_t$ is therefore $\gamma+\psi\,\mathrm{Flow}_t$.

If measured ETF flow activity amplifies the weekend relationship, the interaction coefficient $\psi$ should be positive. This specification tests that prediction. Since contemporaneous Monday flows may respond to the same information as Monday variance, the results do not identify a causal mechanism or establish whether ETF approval contributed to the relationship's emergence.

\subsubsection{Economic magnitude}

The coefficient $\gamma$ measures the change in Monday log variance
associated with a one-unit increase in the weekend factor $W_t$.
To express this relationship in economic terms, we calculate
\begin{equation}
\Delta = \exp\!\left(\hat\gamma\,\sigma_W\right)-1,
\label{eq:magnitude}
\end{equation}
where $\sigma_W$ is the standard deviation of $W_t$ over the
relevant estimation sample. Holding the other regressors fixed,
$100\Delta$ is the approximate percentage change in fitted Monday
variance associated with a one-standard-deviation increase in $W_t$.
Here, fitted variance denotes exponentiated fitted log variance,
and the approximation neglects the small constant $\epsilon$
used in the logarithmic transformation. This quantity refers
to variance rather than volatility.

We also report the incremental coefficient of determination,
$\Delta R^{2}=R^{2}_{\mathrm{HAR}+W}-R^{2}_{\mathrm{HAR}}$,
computed over the same observations. This measures the additional
in-sample explanatory power obtained by adding the weekend factor
to the HAR benchmark.
\subsection{Evaluation and inference}\label{sec:evaluation}

\subsubsection{Forecast evaluation}

Forecast accuracy is assessed using QLIKE and, for the Monday forecasting comparison, RMSE in daily volatility units. Let $\widehat{v}_t$ denote the forecast of $v_t=\ln(RV_t+\epsilon)$, and let $n$ denote the number of forecast evaluation observations. The volatility-scale RMSE is defined as:
\[
\mathrm{RMSE}_{\mathrm{vol}}
=
\left[
\frac{1}{n}\sum_t
\left(\sqrt{h_t}-\sqrt{RV_t}\right)^2
\right]^{1/2},
\]
where $h_t=\exp(\hat v_t)$ is the variance forecast.

The QLIKE loss is
\begin{equation}
\mathrm{QLIKE}(RV_t,h_t)
=
\frac{RV_t}{h_t}
-\ln\left(\frac{RV_t}{h_t}\right)-1,
\label{eq:qlike}
\end{equation}
where $RV_t$ is the cleaned daily variance proxy defined in
Section~\ref{sec:rvmeasure}. We report the average QLIKE loss over the evaluation period. Lower values indicate greater forecast accuracy. Under the conditions described by \citet{patton2011}, QLIKE provides forecast rankings that are robust to noise in the variance proxy.

We compare predictive accuracy using the test of \citet{dm1995}. For losses $\ell_{1,t}$ and $\ell_{2,t}$, define $d_t=\ell_{1,t}-\ell_{2,t}$. The statistic is
\begin{equation}
\mathrm{DM}
=
\frac{\bar d}{\sqrt{\widehat{\mathrm{Var}}(\bar d)}}.
\label{eq:dm}
\end{equation}
For the Monday comparison,
$d_t=\mathrm{QLIKE}_{\mathrm{HAR},t}
-\mathrm{QLIKE}_{\mathrm{HAR}+W,t}$, so positive statistics
favour HAR+$W_t$; $p$-values are one-sided. The baseline variance
estimate is the sample variance of $d_t$ divided by $n$.
A Newey-West estimate with four lags is used as a robustness check.

\subsubsection{Standard errors and inference}

We use heteroskedasticity- and autocorrelation-consistent standard
errors to account for heteroskedasticity and serial dependence
in the predictive regressions. Following \citet{nw1987}, the
estimated long-run covariance matrix is
\begin{equation}
\begin{aligned}
\widehat{\Omega}
&=
\widehat{\Gamma}_0
+\sum_{j=1}^{q}
\left(1-\frac{j}{q+1}\right)
\left(\widehat{\Gamma}_j+\widehat{\Gamma}_j'\right), \\
\widehat{\Gamma}_j
&=
\frac{1}{T}\sum_{t=j+1}^{T}
\hat\eta_t\hat\eta_{t-j}',
\end{aligned}
\label{eq:nw}
\end{equation}
where $\hat\eta_t=\mathbf{x}_t\hat\varepsilon_t$,
$\mathbf{x}_t$ is the vector of regressors, and $T$ is the
number of regression observations. The estimator uses Bartlett
weights with $q=5$ lags. In the Monday-only regressions, these
lags refer to five preceding weekly observations.

Because the three cryptocurrencies share substantial common
variation, the nine crypto-asset pairs cannot be treated as
independent tests. We therefore also report a pooled panel
regression with standard errors clustered by date, allowing
for dependence across pairs observed on the same date.
The pairwise results are interpreted as related evidence
rather than as nine independent confirmations of the channel.

\FloatBarrier

\section{Data}\label{sec:data}

\subsection{Sources and sample}
Daily data are obtained from Yahoo Finance for six assets. The cryptocurrencies are Bitcoin (BTC-USD), Ether
(ETH-USD) and XRP (XRP-USD). The traditional assets are the S\&P~500 index (\^{}GSPC) for equities, COMEX gold
futures (GC=F) for commodities, and the iShares 7-10~Year Treasury Bond ETF (IEF) for fixed income; we refer to
the last as ``Treasury'' throughout, noting that it is a bond fund rather than the on-the-run ten-year note. The
common sample runs from November~2017, bounded by the availability of Ether and XRP histories, to June~2026. Two
auxiliary datasets are used in later tests: the CBOE Volatility Index (\^{}VIX), and daily US spot Bitcoin and Ether
exchange-traded fund net flows obtained from Farside Investors.

The six assets are chosen to make the identification of Section~\ref{sec:weekend} as clean as possible and to
guard against the result being an artefact of any one market. On the cryptocurrency side, Bitcoin and Ether are the
two largest coins by market capitalisation and the two for which US spot funds began trading in January and July~2024 respectively, so they are the natural candidates for any channel that runs through institutional
access. XRP is included as a contrast. It is also among the largest and longest-traded coins, but it had no US spot
fund until November~2025, near the end of the sample, so it allows us to ask whether transmission requires a coin
to be fund-listed. All three have deep and liquid weekend trading and price histories that begin before the
sample, and we stop at three because the major coins load heavily on a single common factor, so adding further
coins would multiply tests without adding independent information.

On the traditional side, each asset represents a distinct asset class and has a genuine weekend closure, which the
identification requires. The S\&P~500 is the benchmark for US equities, the market in which the classical weekend
literature of Section~\ref{sec:lit} was developed, and the one for which an options-implied volatility control,
the VIX, is available. Gold is the reference asset of the cryptocurrency literature, where Bitcoin has repeatedly
been compared with it as a store of value or hedge \citep{dyhrberg2016,baurlucey2010}; its futures reopen on
Sunday evening, giving a shorter closure than equities, a difference we exploit in Section~\ref{sec:reopen}. The
Treasury fund represents fixed income, the asset class through which macroeconomic news is most directly priced.
We use an exchange-traded fund rather than Treasury futures because the fund trades only during the US equity
session and therefore has a full weekend closure, whereas the futures reopen on Sunday evening. Together, the three
span equities, commodities and fixed income, so the result cannot be dismissed as an equity-only artefact.

\subsection{Data preparation}
For each asset $i$ and day $t$ we observe the open $O_{i,t}$, high $H_{i,t}$, low $L_{i,t}$ and close $C_{i,t}$.
Daily logarithmic returns are
\begin{equation*}
r_{i,t} = \ln\!\left(\frac{C_{i,t}}{C_{i,t-1}}\right).
\label{eq:logret}
\end{equation*}
Cryptocurrency bars close at 00:00~UTC and are available on every calendar day, whereas traditional assets trade
only on business days. Rather than force a common calendar by interpolation, which would fabricate observations,
we retain each series on its own trading calendar and align them only where a test requires it. The variance series are cleaned as described in Section~\ref{sec:rvmeasure}. Zero variance estimates are treated as missing because their logarithms are undefined. Where a Monday is a US market holiday, the traditional asset has no
observation and that week is dropped from the main weekend tests; a separate placebo test maps the weekend to the
first subsequent trading day. The post-ETF joint and VIX-regime specifications contain 115 Monday observations for the S\&P~500 and Treasury (IEF), and 116 for gold. Adding Friday VIX does not reduce these estimation samples.

\subsection{Descriptive statistics}

Table~\ref{tab:desc} reports summary statistics of daily
logarithmic returns over the common sample period.
Annualised return volatility is calculated by multiplying
the daily return standard deviation by $\sqrt{365}$ for
cryptocurrencies and $\sqrt{252}$ for traditional assets. Mean daily returns are positive for all six assets.
Cryptocurrencies exhibit substantially greater return
variability than the traditional assets, with daily standard
deviations ranging from $3.49\%$ for Bitcoin to $5.58\%$ for
XRP. Among the traditional assets, daily standard deviations
range from $0.43\%$ for Treasury to $1.21\%$ for the S\&P~500.
XRP has the highest annualised return volatility at $106.6\%$,
while Treasury has the lowest at $6.9\%$.

Bitcoin, Ether, the S\&P~500 and gold exhibit negative
skewness, whereas XRP exhibits positive skewness and
Treasury returns are approximately symmetric. All six
assets have positive excess kurtosis, indicating heavier
tails than a normal distribution. These statistics describe
unconditional daily returns; they do not establish the
weekend predictive relationship examined in
Section~\ref{sec:weekend}.

These distributional differences have implications for risk
measurement and modelling. The greater return variability
of cryptocurrencies indicates larger fluctuations at the
daily horizon, while negative skewness in several assets
highlights downside asymmetry. Positive excess kurtosis
across all six assets suggests that a normal distribution
may underrepresent extreme returns. These features motivate
the use of robust inference and complementary forecast
evaluation criteria, although they do not themselves provide
evidence of a weekend volatility channel.

\begin{table}[H]\centering
\caption{Summary statistics of daily logarithmic returns, common sample (November~2017 to June~2026). Annualised
volatility uses $\sqrt{365}$ for cryptocurrencies and $\sqrt{252}$ for traditional assets.}
\label{tab:desc}
\begin{tabular}{lccccc}
\toprule
Asset & Mean (\%) & Std.\ dev.\ (\%) & Skewness & Excess kurtosis & Ann.\ vol.\ (\%)\\
\midrule
Bitcoin        & 0.074 & 3.49 & $-0.74$ & 13.6 & 66.6\\
Ether          & 0.054 & 4.50 & $-0.77$ & 10.6 & 85.9\\
XRP            & 0.054 & 5.58 & $1.12$  & 20.6 & 106.6\\
S\&P 500       & 0.049 & 1.21 & $-0.66$ & 15.0 & 19.2\\
Gold           & 0.057 & 1.09 & $-0.90$ & 10.4 & 17.2\\
Treasury (IEF) & 0.004 & 0.43 & $0.16$  & 3.1  & 6.9\\
\bottomrule
\end{tabular}
\end{table}

\begin{figure}[H]
\centering
\includegraphics[width=\textwidth]{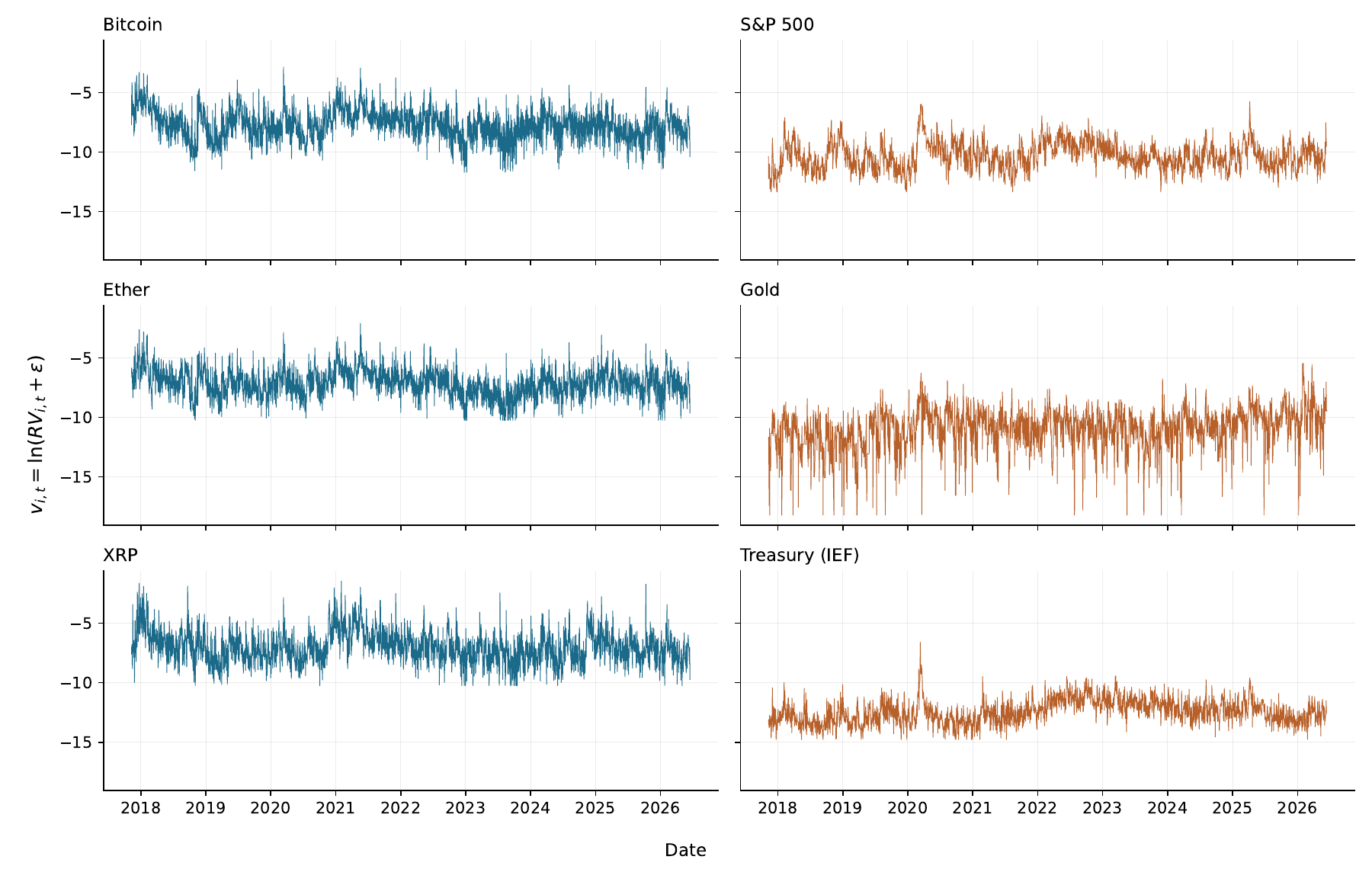}
\caption{Daily log variance for the six assets.
Each panel plots $v_{i,t}=\ln(RV_{i,t}+\epsilon)$,
using the Garman-Klass variance proxy defined in
Section~\ref{sec:rvmeasure}.}
\label{fig:logvariance}
\end{figure}

Figure~\ref{fig:logvariance} shows daily log variance across the six assets. Cryptocurrency variance generally exceeds that of the traditional assets, while Bitcoin and Ether display broadly similar temporal patterns. Treasury (IEF) variance is visibly elevated around 2022–2023 relative to much of the earlier sample. Gold exhibits pronounced downward spikes associated with very low estimated within-session variance; the most extreme observations are bounded by the asset-specific lower winsorisation threshold. These descriptive patterns do not, by themselves, establish cross-market transmission.

\section{Results}\label{sec:results}
We first examine the emergence of the weekend predictive relationship and assess whether weekend cryptocurrency variance provides information beyond Friday activity and the traditional assets' own variance histories. We then
distinguish common cryptocurrency-market variation from coin-specific information, quantify the economic magnitude of the relationship and evaluate its out-of-sample forecasting value. Next, we assess robustness and examine variation across weekdays, holiday reopening dates and VIX regimes, and we investigate whether ETF-flow activity helps explain the relationship. Finally, complementary comparisons with flexible forecasting models and aggregate connectedness measures conclude the section.

\subsection{The channel and its emergence}
\label{sec:emerge}

Table~\ref{tab:emerge} reports the average weekend coefficient across nine separate crypto–asset regressions within disjoint sub-periods. Each regression relates an individual cryptocurrency’s weekend variance to Monday variance in one traditional asset, controlling for that cryptocurrency’s Friday variance and the traditional asset’s daily, weekly and monthly HAR terms. The COVID–Recovery sub-period begins on 20 February 2020, and the post-ETF sub-period begins on 11 January 2024. The reported mean is the simple average of the nine estimated weekend coefficients, rather than a pooled panel estimate.

\begin{table}[H]
\centering
\caption{Emergence of the weekend channel: average of nine
pairwise weekend coefficients by sub-period.}
\label{tab:emerge}
\begin{tabular}{lcc}
\toprule
Sub-period & Mean $\hat\gamma$
& Positive and significant pairs ($p<0.05$)\\
\midrule
Pre-COVID (2017-11 to 2020-02)
& $-0.03$ & 0/9\\
COVID-Recovery (2020-02 to 2021-12)
& 0.03 & 0/9\\
LUNA/FTX era (2022-01 to 2023-12)
& 0.14 & 2/9\\
Post-ETF (2024-01 to 2026-06)
& 0.35 & 9/9\\
\bottomrule
\end{tabular}
\end{table}

\begin{figure}[H]\centering
\includegraphics[width=\textwidth]{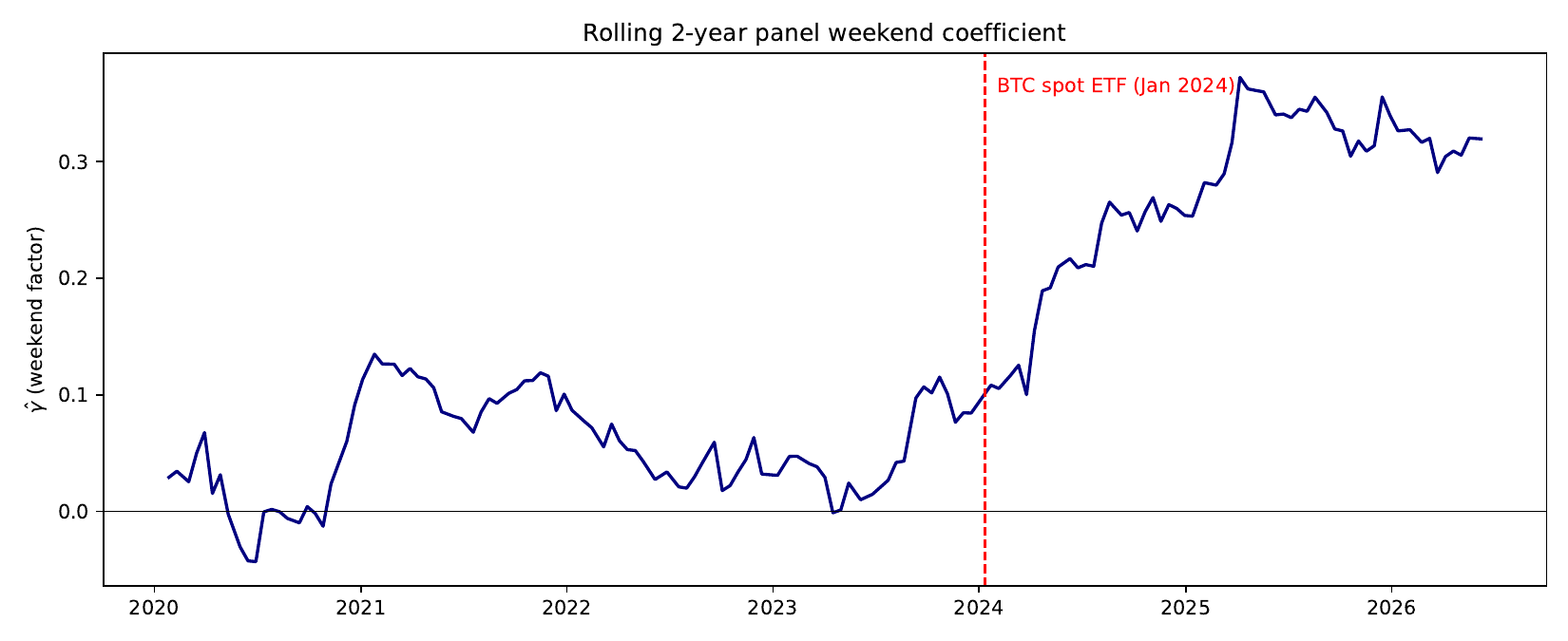}
\caption{Rolling two-year panel estimate of the weekend coefficient $\hat\gamma$. The dashed line marks the
January~2024 US spot-Bitcoin ETF approvals.}
\label{fig:rollgamma}
\end{figure}

The mean coefficient is close to zero and slightly negative
before 2020, and remains relatively weak during 2022-2023,
when only two of the nine pairwise coefficients are positive
and significant at the $5\%$ level. After January~2024,
the mean rises to $0.35$, with all nine coefficients positive
and significant. Figure~\ref{fig:rollgamma} presents rolling
two-year panel estimates. The coefficient remains close
to zero over much of 2020-2023, increases from late 2023
and subsequently settles around $0.30$ during 2025 and 2026.

We also estimate the break date without imposing the ETF approval date. The sup-Wald procedure in \eqref{eq:supwald} places the break in October~2023,
with $\sup\text{-}\mathcal{W}\approx8.8$, corresponding to a maximum absolute test statistic of $2.97$. Comparison with the critical values of \citet{andrews1993} indicates significance at the $10\%$ level and borderline
evidence at the $5\%$ level. Excluding the week of 16~October~2023 moves the estimated break to early November~2023.

The estimated break precedes January~2024 and is consistent with the increase in the rolling coefficient during late 2023. This timing is compatible with anticipation of ETF approval, but does not establish that anticipation caused
the change. January~2024 therefore provides a subsequent comparison date rather than the estimated onset of the relationship. These results suggest that the weekend predictive relationship strengthened across the crypto-asset pairs after January~2024, rather than being confined to a single pair. However, the pairwise estimates are not independent, and the timing alone does not establish that ETF approval caused the change.

\subsection{The weekend volatility channel}

We examine two aspects of the weekend predictive relationship. First, we assess whether weekend cryptocurrency variance provides additional information about Monday variance beyond Friday cryptocurrency activity and the traditional asset's own variance history. Second, we examine whether this
predictive information reflects variation shared across cryptocurrency markets or coin-specific components.

\subsubsection{Identification: weekend versus Friday}
The weekend and Friday factors are correlated ($\rho=0.67$).
Table~\ref{tab:coll} therefore reports the joint, weekend-only
and Friday-only specifications described in
Section~\ref{sec:weekend}. The weekend coefficient is positive
and significant in both the joint and weekend-only models.
In the weekend-only specification, it is $0.256$ for the
S\&P~500, $0.447$ for gold and $0.246$ for Treasury,
with $p\leq0.005$ in all three cases. The Friday-only coefficients are close to zero and statistically insignificant, with $p$-values ranging from $0.521$ to $0.883$. Separately, the one-sided contrast in \eqref{eq:wald} rejects in favour of the weekend coefficient in all nine coin-asset pairwise regressions.

Together, these findings support additional predictive
content associated with weekend cryptocurrency variance.
However, the difference between a significant weekend
coefficient and an insignificant Friday coefficient does
not itself establish that the coefficients differ; this
comparison is assessed by the one-sided contrast.
The results provide predictive evidence rather than
causal identification of transmission during market closure.

\begin{table}[H]\centering
\caption{Weekend versus Friday: joint, weekend-only and Friday-only specifications, post-2024.}
\label{tab:coll}
\begin{tabular}{lcccccc}
\toprule
 & \multicolumn{2}{c}{Joint ($\gamma$)} & \multicolumn{2}{c}{Weekend-only ($\gamma$)} & \multicolumn{2}{c}{Friday-only ($\delta$)}\\
\cmidrule(lr){2-3}\cmidrule(lr){4-5}\cmidrule(lr){6-7}
Asset & $\hat\gamma$ & $p$ & $\hat\gamma$ & $p$ & $\hat\delta$ & $p$\\
\midrule
S\&P 500       & 0.349 & 0.001 & 0.256 & 0.005 & $-0.015$ & 0.869\\
Gold           & 0.645 & <0.001 & 0.447 & 0.001 & $-0.017$ & 0.883\\
Treasury (IEF) & 0.340 & 0.001 & 0.246 & 0.001 & 0.034 & 0.521\\
\bottomrule
\end{tabular}
\end{table}
\par\smallskip
\begin{minipage}{\textwidth}
\footnotesize
\textit{Notes:} All specifications include the HAR controls.
Reported coefficient $p$-values use Newey-West standard
errors with five lags.
\end{minipage}

\subsubsection{Market-level, not coin-specific}

Table~\ref{tab:orth} reports the results of adding each coin's residual from equation~(\ref{eq:orth}) separately to
the joint specification. The common factor remains positive and significant in all nine specifications, with
coefficients between 0.33 and 0.64 and $p \le 0.002$. None of the nine coin-specific residuals is significant
at the 5\% level; two are significant at the 10\% level (Bitcoin for gold, $p = 0.081$, and XRP for Treasury,
$p = 0.066$), with opposite signs. The predictive content is therefore associated mainly with variation shared
across cryptocurrencies rather than with any individual coin.

\begin{table}[H]\centering
\caption{Common crypto factor versus coin-specific components, post-2024. Each row adds one coin's
residual $\hat u_{c,t}$ from equation~(\ref{eq:orth}) to the joint specification~(\ref{eq:weekend}), which already contains $W_t$,
$F_t$ and the HAR terms. Newey-West standard errors with five lags.}
\label{tab:orth}
\begin{tabular}{llcccc}
\toprule
 & & \multicolumn{2}{c}{Common factor $W_t$} & \multicolumn{2}{c}{Coin residual $\hat u_{c,t}$}\\
\cmidrule(lr){3-4}\cmidrule(lr){5-6}
Asset & Residual added & $\hat\gamma$ & $p$ & Coefficient & $p$\\
\midrule
S\&P 500       & Bitcoin & 0.344 & 0.001   & $-0.028$ & 0.858\\
               & Ether   & 0.349 & 0.001   & $0.013$  & 0.931\\
               & XRP     & 0.347 & 0.001   & $0.012$  & 0.901\\
\addlinespace
Gold           & Bitcoin & 0.592 & 0.002   & $-0.372$ & 0.081\\
               & Ether   & 0.639 & $<$0.001 & $0.115$ & 0.697\\
               & XRP     & 0.627 & 0.001   & $0.200$  & 0.221\\
\addlinespace
Treasury (IEF) & Bitcoin & 0.358 & $<$0.001 & $0.152$ & 0.289\\
               & Ether   & 0.328 & 0.001   & $0.173$  & 0.308\\
               & XRP     & 0.349 & $<$0.001 & $-0.187$ & 0.066\\
\bottomrule
\end{tabular}
\end{table}

These results are consistent with predictive content arising
primarily from variation shared across cryptocurrency markets.
This may also explain the similar pairwise findings across
Bitcoin, Ether and XRP. The coin-specific components provide
no statistically significant additional predictive information
beyond the common factor and the other controls. However,
these findings do not rule out ETF-related explanations,
since ETF activity may also affect the common market factor.
The underlying mechanism remains unresolved.

\begin{figure}[H]\centering
\includegraphics[width=\textwidth]{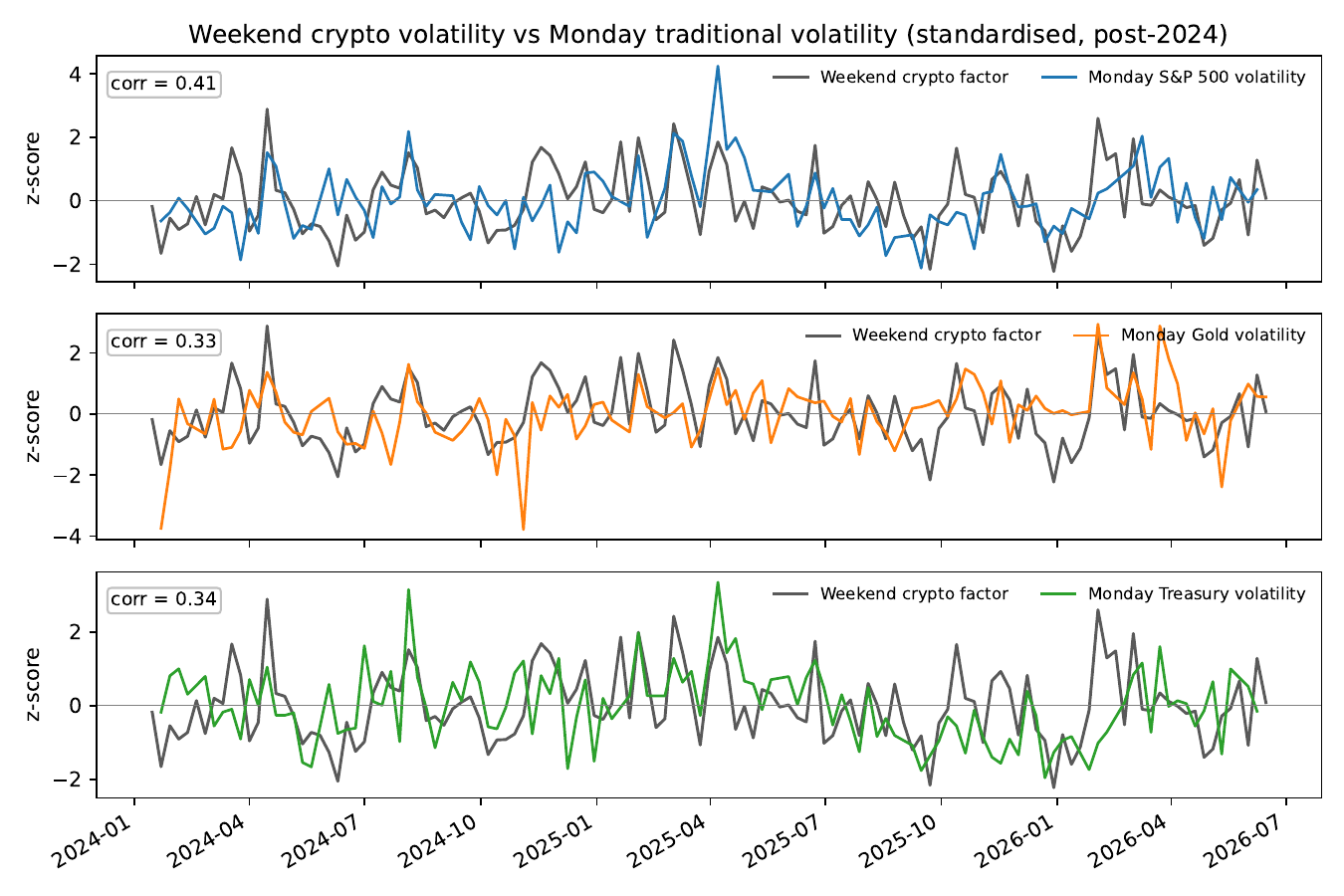}
\caption{Weekend cryptocurrency volatility factor and Monday traditional-market realised volatility
(standardised), post-2024. Reported correlations are unconditional, that is, before the HAR controls of
equation~\eqref{eq:weekend}.}
\label{fig:wkmon}
\end{figure}

Furthermore, Figure~\ref{fig:wkmon} illustrates this common-market relationship by plotting the standardised weekend factor alongside Monday volatility for each traditional asset\footnote{For comparison, the aggregate connectedness analysis is reported in Section~\ref{sec:conn}.}. The unconditional correlations are 0.41 for the S\&P~500, 0.33 for gold and 0.34 for Treasury. These positive associations complement Table~\ref{tab:orth}, which assesses the factor's predictive content conditional on the HAR controls, Friday cryptocurrency variance and coin-specific components.

\subsection{Economic magnitude and out-of-sample value}

We assess the practical relevance of the weekend predictive relationship in two ways. First, we translate the estimated weekend coefficient into the percentage change in fitted Monday variance associated with a one-standard-deviation increase in the weekend factor. Second, we evaluate whether
adding this factor to the HAR benchmark improves out-of-sample Monday variance forecasts.

\subsubsection{Economic magnitude}
Table~\ref{tab:mag} translates the weekend-only estimates into economic terms using \eqref{eq:magnitude}. With $\sigma_W=0.97$ over the post-2024 sample, a one-standard-deviation increase in the weekend factor $W_t$ is associated with approximately $28\%$ higher fitted Monday variance for the S\&P~500, $54\%$ for gold and $27\%$ for Treasury, holding the HAR controls fixed. Adding $W_t$ increases the in-sample $R^2$ by between $3.9$ and $7.0$ percentage points relative to the HAR benchmark. These estimates indicate a substantial
relationship between weekend cryptocurrency variance and Monday variance beyond the traditional asset's own variance history.

\begin{table}[H]\centering
\caption{Economic magnitude of the weekend factor, post-2024. The one-standard-deviation effect is
$\exp(\hat\gamma\,\sigma_W)-1$, using the weekend-only coefficients of Table~\ref{tab:coll}.}
\label{tab:mag}
\begin{tabular}{lcccc}
\toprule
Asset & Weekend-only $\hat\gamma$ & $\sigma_W$ & 1-SD effect on Monday variance & $\Delta R^2$ (pp)\\
\midrule
S\&P 500       & 0.256 & 0.97 & $+28\%$ & 3.9\\
Gold           & 0.447 & 0.97 & $+54\%$ & 6.2\\
Treasury (IEF) & 0.246 & 0.97 & $+27\%$ & 7.0\\
\bottomrule
\end{tabular}
\end{table}

\subsubsection{Out-of-sample value}
The in-sample findings are complemented by the out-of-sample
forecast comparison in Table~\ref{tab:oos}. The evaluation
covers 103-104 Monday observations per asset from
April~2024 to June~2026, with models re-estimated weekly
using an expanding window. Adding $W_t$ to the HAR
benchmark reduces QLIKE by approximately $22\%$ for the
S\&P~500, $17\%$ for Treasury and $11\%$ for gold.
RMSE, measured in daily volatility units, also decreases
for all three assets.

The one-sided Diebold-Mariano test uses QLIKE loss
differences, with positive statistics favouring HAR$+W_t$.
The improvement is significant at the $5\%$ level for
Treasury ($p=0.042$), near the $5\%$ threshold for gold
($p=0.050$), and significant at the $10\%$ level for
the S\&P~500 ($p=0.066$). Thus, the weekend factor
lowers both forecast loss measures across all three
assets, although the statistical evidence varies
across markets.

\begin{table}[H]
\centering
\caption{Out-of-sample Monday volatility forecasts:
HAR versus HAR$+W_t$, April~2024-June~2026.}
\label{tab:oos}
\small
\begin{tabular}{lcccccc}
\toprule
Asset & QLIKE (HAR) & QLIKE (HAR$+W_t$)
& RMSE (HAR) & RMSE (HAR$+W_t$) & DM & $p$\\
\midrule
S\&P 500
& 0.695 & 0.545 & 0.00547 & 0.00514 & 1.51 & 0.066\\
Gold
& 3.182 & 2.845 & 0.00878 & 0.00836 & 1.64 & 0.050\\
Treasury (IEF)
& 0.424 & 0.350 & 0.00104 & 0.00098 & 1.73 & 0.042\\
\bottomrule
\end{tabular}
\end{table}

\begin{figure}[H]\centering
\includegraphics[width=\textwidth]{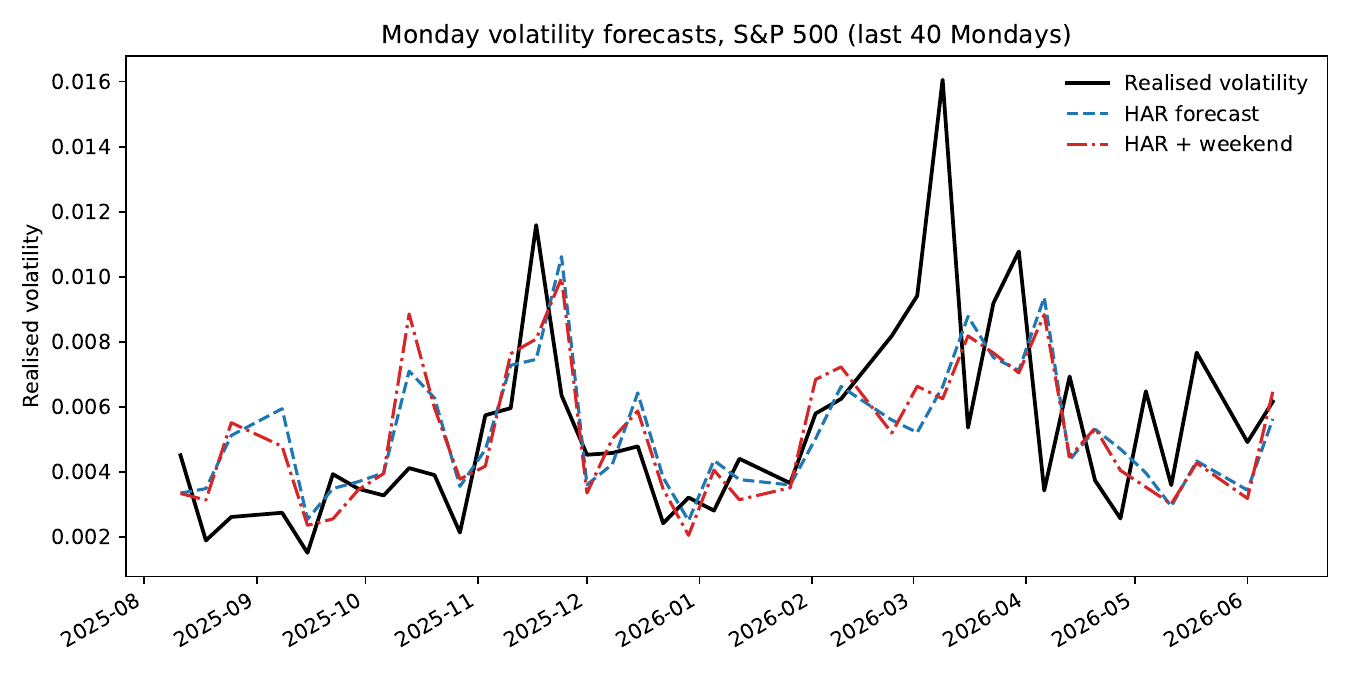}
\caption{Monday realised volatility for the S\&P~500 against the HAR and HAR-plus-weekend forecasts over the last
40 Mondays of the sample. Both forecast series are one-step-ahead, walk-forward forecasts.}
\label{fig:fcsp}
\end{figure}

Furthermore, Figure~\ref{fig:fcsp} illustrates the forecast
comparison for the S\&P~500 over the final forty Mondays
of the sample. The HAR and HAR$+W_t$ forecasts are visually
similar, and both underestimate some of the largest
realised-volatility spikes. The lower forecast losses
reported in Table~\ref{tab:oos} therefore reflect improvements
that are not always apparent from the plotted paths.

These findings complement the benchmark comparison in Section~\ref{sec:bench}. The flexible models did not consistently outperform HAR, whereas adding the weekend factor reduces forecast losses for all three traditional assets. Within the specifications examined, the results support the value of incorporating weekend information alongside the asset's own variance history.

\subsection{Robustness}
We assess the robustness of the weekend predictive relationship using additional controls, alternative inference procedures and influential-observation checks. We then examine whether the relationship is concentrated
on Mondays or also appears on other weekdays and holiday-delayed reopening dates. Finally, we investigate whether its strength varies with the preceding Friday's VIX. 

\subsubsection{Additional controls and sample-sensitivity checks}

Three checks assess the sensitivity of the weekend predictive relationship to pre-weekend market uncertainty, dependence across crypto-asset pairs observed on the same date, and influential observations. Table~\ref{tab:robust} summarises these checks alongside the other robustness analyses.

\begin{table}[H]
\centering
\caption{Robustness checks on the weekend channel.
Results refer to the post-2024 sample; ``pairs'' denotes
the nine crypto-asset combinations and ``assets'' the
three traditional assets.}
\label{tab:robust}
\small
\begin{tabular}{p{4.3cm}p{4.6cm}p{4.6cm}}
\toprule
Check & Concern addressed & Result \\
\midrule
Weekend-only specification
& Sensitivity to joint inclusion of $W_t$ and $F_t$
($\rho=0.67$)
& Significant for 3/3 assets, $p\leq0.005$ \\
\addlinespace

Wald test of $H_0:\gamma\leq\delta$
& Whether the weekend coefficient exceeds the Friday
coefficient
& Null rejected in favour of $\gamma>\delta$ for 9/9
pairs \\
\addlinespace

Friday VIX control
& Whether pre-weekend equity-market uncertainty explains
the weekend relationship
& 8/9 pairs remain significant \\
\addlinespace

Date-clustered panel
& Dependence across pairs observed on the same date
& $\hat{\gamma}=0.38$, $t=5.6$, 116 Mondays \\
\addlinespace

Most influential Monday removed
& Sensitivity to a single influential observation
& 8/9 pairs remain significant \\
\addlinespace

Orthogonalisation
& Common versus coin-specific predictive information
& Common factor significant for all three assets;
no coin-specific residual coefficient significant \\
\addlinespace

Endogenous break (sup-Wald)
& Sensitivity to imposing the break date
& October~2023, $\sup|t|=2.97$
($\sup\text{-}\mathcal{W}\approx8.8$) \\
\addlinespace

Weekday comparison (Tue, Wed)
& Whether the relationship is concentrated on Mondays
& Monday coefficient largest for 3/3 assets;
Monday-Wednesday contrast significant at 5\%
for Treasury \\
\addlinespace

Holiday-week reopen test
& Monday concentration versus a generic reopening
relationship
& Negative holiday interactions; only 11 holiday
weeks \\
\addlinespace

Walk-forward out-of-sample
& Whether forecast gains extend beyond the estimation
sample
& QLIKE falls by 11-22\%; one-sided DM
$p=0.042$ for Treasury, $0.050$ for gold and
$0.066$ for the S\&P~500 \\
\addlinespace

Alternative range estimators (\ref{appendix}) & Dependence on the Garman-Klass estimator & Weekend effect significant for 3/3 assets under Parkinson and Rogers-Satchell; break date sensitive\\
\bottomrule
\end{tabular}
\end{table}

The first check adds the preceding Friday's VIX to the pairwise regressions. This assesses whether the weekend coefficient reflects equity-market uncertainty already observable before the weekend. Eight of the nine weekend coefficients remain statistically significant, with little change in their magnitudes. The relationship is therefore not fully explained by pre-weekend
uncertainty as measured by VIX, although other sources of uncertainty may remain relevant.

The second check addresses dependence across the nine crypto-asset pairs. Since the cryptocurrencies share substantial common variation, the pairwise findings should not be interpreted as nine independent confirmations. Pooling the pairs in a panel with standard errors clustered by date gives a weekend
coefficient of $0.38$ and a $t$-statistic of $5.6$ across 116 Mondays. The relationship remains significant after allowing for dependence across
pairs within each Monday. However, date clustering does not address serial dependence across Mondays.

The third check examines sensitivity to influential observations. Each pairwise regression is re-estimated after removing its most influential post-2024 Monday. Eight of the nine weekend coefficients remain statistically significant, suggesting that the pairwise findings are generally robust to removing a single influential observation.

Together, these checks support the robustness of the weekend predictive relationship to the Friday VIX control, within-date dependence and individual
influential observations.

\subsubsection{Concentration on Mondays}\label{sec:reopen}
Table~\ref{tab:reopen} reports the weekend factor's coefficient
for Monday, Tuesday and Wednesday sessions. The coefficient
is largest on Monday in all three markets. The formal
Monday-minus-Wednesday contrast is significant for Treasury
(one-sided $p=0.02$), significant at the $10\%$ level for
the S\&P~500 ($p=0.07$), and insignificant for gold.
Gold's weaker evidence of Monday concentration is compatible
with its shorter closure, since its futures reopen on
Sunday evening. However, the comparison does not establish
that closure length explains the differences across assets.

\begin{table}[H]\centering
\caption{Reopen gradient: weekend factor coefficient by weekday, post-2024.}
\label{tab:reopen}
\begin{tabular}{lcccccc}
\toprule
 & \multicolumn{2}{c}{Monday} & \multicolumn{2}{c}{Tuesday} & \multicolumn{2}{c}{Wednesday}\\
\cmidrule(lr){2-3}\cmidrule(lr){4-5}\cmidrule(lr){6-7}
Asset & $\hat\beta$ & $p$ & $\hat\beta$ & $p$ & $\hat\beta$ & $p$\\
\midrule
S\&P 500       & 0.285 & <0.001 & 0.170 & 0.060 & 0.137 & 0.011\\
Gold           & 0.432 & 0.002 & $-0.195$ & 0.321 & 0.259 & 0.043\\
Treasury (IEF) & 0.252 & 0.001 & 0.077 & 0.338 & 0.077 & 0.353\\
\bottomrule
\end{tabular}
\end{table}

\textit{Notes:} Separate regressions are estimated for each
weekday using available post-2024 observations. Each regression
includes an intercept, the preceding trading session's log
variance and the weekend factor for the same calendar week.
Unlike Table~\ref{tab:coll}, weekly and monthly HAR controls are omitted.
Samples are not restricted to common weeks. Coefficient
$p$-values use Newey-West standard errors with five lags.

We also examine the eleven weeks in which a Monday holiday
moves the reopening session to Tuesday. The holiday
interaction is negative and significant for the S\&P~500
($-0.45$, $p=0.01$) and Treasury ($-0.50$, $p=0.01$),
and negative but insignificant for gold ($-0.23$, $p=0.36$). These estimates indicate a weaker weekend relationship on holiday-delayed reopening dates. Whether the relationship remains significant on those dates requires a test of the baseline weekend coefficient plus the holiday interaction.

Given the small number of holiday weeks, these findings
are interpreted cautiously. They suggest that the
relationship may be concentrated on Mondays rather than
on the first trading session after any weekend closure.
This limits the reopening interpretation, while leaving
the main Monday predictive results as a separate finding.

\subsubsection{VIX-dependent variation in the weekend relationship}
\label{sec:vix_regimes}

We examine whether the weekend cryptocurrency-Monday variance
relationship varies with pre-weekend equity-market uncertainty.
Using the post-ETF sample beginning on 11 January 2024, we
classify Monday observations according to the preceding Friday's
VIX. Where Friday's VIX is missing, we use the latest earlier
available observation. Two thresholds are considered: the median
and the 75th percentile of log Friday VIX in each asset's
estimation sample. Observations strictly above the relevant
threshold form the high-VIX regime; the remaining observations
form the low-VIX regime. Under the 75th-percentile definition,
the comparison is therefore between the upper quarter and the
lower three quarters of the distribution.

We extend the joint weekend-Friday specification in equation~\ref{eq:weekend} by allowing its intercept and all slopes to vary across
these regimes:
\begin{equation*}
\begin{aligned}
v_t^{(m)}
={}&
\sum_{r\in\{L,H\}} I_{m,t}^{(r)}
\Big[
\alpha_r
+\beta_{d,r}v_{t-1}^{(m)}
+\beta_{w,r}\bar v_{t-1:t-5}^{(m)}
+\beta_{mo,r}\bar v_{t-1:t-22}^{(m)}
\\
&\hspace{4cm}
+\gamma_r W_t+\delta_r F_t
\Big]
+\varepsilon_t .
\end{aligned}
\label{eq:vix_regime}
\end{equation*}

Here, $I_{m,t}^{(r)}$ is an indicator identifying the VIX regime
for traditional asset $m$ on Monday $t$. Specifically,
\[
I_{m,t}^{(H)}
= \mathbf{1}\!\left\{\ln(\mathrm{VIX}_{\mathrm{Fri}(t)})>c_m\right\},
\qquad
I_{m,t}^{(L)}=1-I_{m,t}^{(H)},
\]
where $\mathbf{1}\{\cdot\}$ equals one when the stated condition
holds and zero otherwise, $\mathrm{VIX}_{\mathrm{Fri}(t)}$ is the
preceding Friday's VIX (or the latest earlier available observation
if Friday's value is missing), and $c_m$ is the median or 75th
percentile of log Friday VIX in asset $m$'s post-ETF estimation
sample, depending on the specification. Observations equal to
the threshold are assigned to the low-VIX regime.
The model is estimated separately for each traditional asset and threshold definition; asset and threshold indices on the coefficients are suppressed for simplicity. The HAR terms retain their definitions from Section~\ref{sec:har} and are constructed from the asset's daily trading history, rather than from Monday observations alone. The weekend and Friday cryptocurrency
factors, $W_t$ and $F_t$, retain their definitions in
equation~\ref{eq:wkfactor}.

Each asset-specific threshold defines a separate regression specification. VIX is used to assign observations to regimes; it does not enter as a separate explanatory variable.\footnote{For the S\&P~500, the median split contains 58 lower-VIX and 57 higher-VIX Mondays. The 75th-percentile split contains 86 lower-VIX and 29 higher-VIX Mondays.} Newey-West standard errors with five lags are used to construct 95\% confidence intervals. A Wald test of $H_0:\gamma_H=\gamma_L$ assesses whether the weekend coefficient differs between regimes.

\begin{figure}[H]
    \centering
    \includegraphics[width=0.85\linewidth]{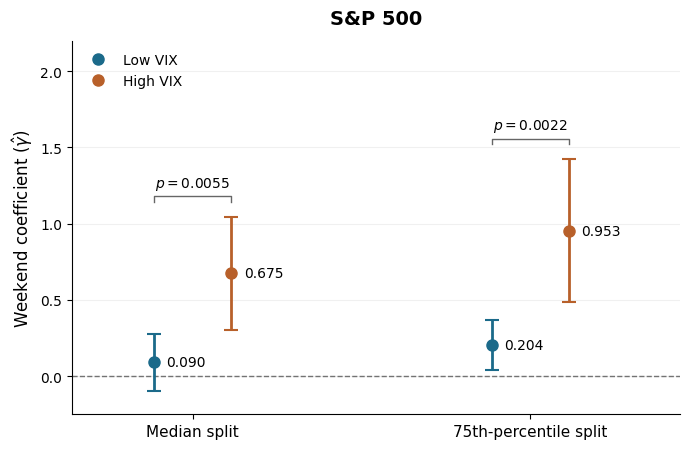}
    \caption{S\&P~500 weekend coefficients by Friday VIX regime.
    Bars show 95\% confidence intervals using Newey-West
    standard errors with five lags. Intercepts and all slopes
    vary across regimes. Difference $p$-values are Wald tests
    of equal weekend coefficients, unadjusted for multiple
    comparisons. Thresholds are asset-specific.}
    \label{fig:vix_sp500}
\end{figure}

\begin{figure}[H]
    \centering
    \includegraphics[width=0.85\linewidth]{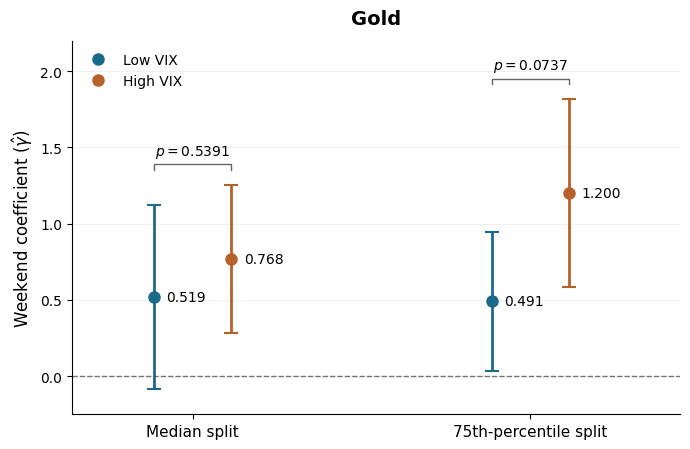}
    \caption{Gold weekend coefficients by Friday VIX regime.
    Estimation and inference follow
    Figure~\ref{fig:vix_sp500}.}
    \label{fig:vix_gold}
\end{figure}

\begin{figure}[H]
    \centering
    \includegraphics[width=0.85\linewidth]{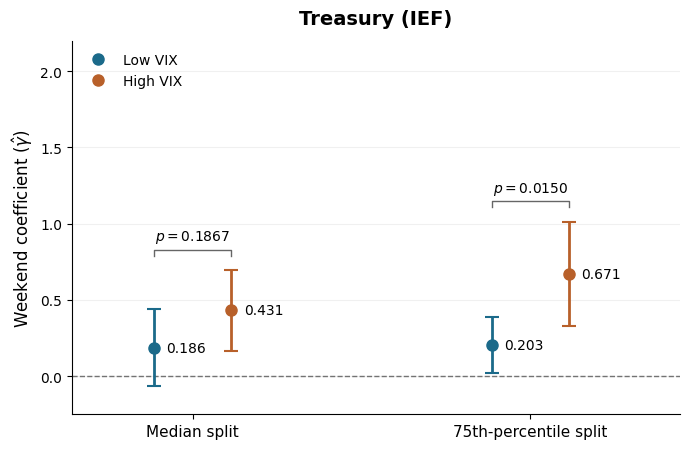}
    \caption{Treasury ETF (IEF) weekend coefficients by Friday
    VIX regime. Estimation and inference follow
    Figure~\ref{fig:vix_sp500}.}
    \label{fig:vix_ief}
\end{figure}

Figures~\ref{fig:vix_sp500}-\ref{fig:vix_ief} show larger weekend coefficients in the high-VIX regime for all three assets, although the evidence of regime differences varies. For the S\&P~500, the coefficient increases from $0.090$ to $0.675$ under the median split ($p=0.0055$), and from
$0.204$ to $0.953$ under the 75th-percentile split ($p=0.0022$). Thus, the association between weekend cryptocurrency variance and Monday S\&P~500 variance is stronger when pre-weekend VIX is elevated under both threshold definitions.

The evidence is less consistent for gold and IEF. For gold, the low- and high-VIX coefficients are $0.519$ and $0.768$ under the median split ($p=0.5391$), and $0.491$ and $1.200$ under the 75th-percentile split ($p=0.0737$). Neither
difference is significant at the 5\% level. For IEF, the median-split difference is also insignificant ($0.186$ versus $0.431$; $p=0.1867$), whereas the 75th-percentile split produces a significant difference ($0.203$ versus $0.671$; $p=0.0150$).

Overall, VIX-dependent variation is most clearly supported for the S\&P~500, while the IEF finding depends on the threshold and the gold evidence remains weaker. VIX represents equity-market uncertainty rather than gold- or Treasury-specific implied volatility. The reported $p$-values are unadjusted for the six comparisons, and the two threshold specifications use overlapping observations. These findings indicate that the weekend predictive relationship
is stronger during periods of elevated pre-weekend VIX, with the clearest evidence for the S\&P~500.

\subsection{The exchange-traded-fund flow mechanism}
We examine whether the weekend relationship is stronger when institutional ETF flow activity is greater. In \eqref{eq:flow}, the interaction between $W_t$ and
total Bitcoin and Ether spot-ETF flow activity is statistically insignificant for the S\&P~500 ($p=0.30$) and gold ($p=0.85$), and significantly negative for
Treasury ($t=-2.6$). These results do not support the hypothesis that the weekend relationship strengthens with greater measured ETF flow activity.

The insignificant interactions do not rule out ETF-related transmission, since the flow proxy may not capture all relevant institutional activity. Similarly, the negative Treasury interaction indicates a weaker association between $W_t$ and Monday log variance when flow activity is greater, but does not identify the reason for this difference. Broader links through overlapping holders, dealers and risk-management systems remain possible explanations. However, these mechanisms are not directly tested here, and contemporaneous Monday flows may respond to the same information as Monday variance. The underlying mechanism therefore remains unresolved.

\subsection{Complementary forecasting and connectedness analyses}
The following analyses place the weekend results in a broader context by comparing HAR with flexible forecasting models and examining aggregate daily
volatility connectedness.

\subsubsection{Comparison with flexible forecasting models}\label{sec:bench}

We first assess whether flexible forecasting models improve
on the HAR benchmark. Machine learning has delivered economic
gains in empirical asset pricing \citep{gukellyxiu2020}, while
neural-network and high-dimensional methods have improved
realised-volatility forecasts in some settings
\citep{bucci2020,rahimikia2020}. Using the expanding-window
walk-forward design in Section~\ref{sec:walkforward}, we compare
HAR with a random forest \citep{breiman2001}, gradient boosting
\citep{friedman2001,chenguestrin2016} and a long short-term
memory network \citep{hochreiter1997}. Each flexible model receives the HAR predictors and the preceding day's absolute return. 

Hyperparameters are tuned separately for each asset through an exhaustive grid search evaluated using QLIKE under time-series cross-validation within the first $60\%$ of each series. This produces $372$ fitted configurations in total. Forecast evaluation begins at observation $400$, which falls within the tuning block, creating a partial overlap between the tuning and evaluation periods. Consequently, the evaluation is not strictly out of sample with respect to hyperparameter selection and may favour the machine-learning models. Nevertheless, these models do not consistently outperform HAR:
gradient boosting achieves lower QLIKE for XRP, while the random
forest achieves lower QLIKE for gold. HAR and the tree-based models are re-estimated every $21$ observations, whereas the LSTM is re-estimated every $63$ observations. The network’s inputs and target are standardised using training-window statistics only.

The HAR benchmark achieves the lowest reported QLIKE for
four of the six assets: Bitcoin ($0.763$), Ether ($0.640$,
tied with the random forest), the S\&P~500 ($0.454$) and
Treasury ($0.402$). Gradient boosting performs better for
XRP ($1.004$ compared with $1.032$ for HAR), while the
random forest performs better for gold ($2.093$ compared
with $2.370$). The recurrent network does not achieve the
lowest QLIKE for any asset. Tuning improves the tree models in eleven of the twelve asset-model combinations relative to their default settings, but does not change the overall ranking. For Treasury, the random forest and gradient boosting losses remain
approximately $40\%$ and $25\%$ higher than the HAR loss,
respectively. The results therefore show that greater
model flexibility does not consistently improve forecasting
performance in this sample.

These findings support the use of HAR as a parsimonious
benchmark for the weekend-channel analysis.
Any forecasting improvement from adding $W_t$ is assessed
relative to this benchmark. The comparison does not rule
out gains from other flexible models or from models that
explicitly include weekend information.

\subsubsection{Aggregate volatility connectedness}\label{sec:conn}
\begin{figure}[H]\centering
\includegraphics[width=0.82\textwidth]{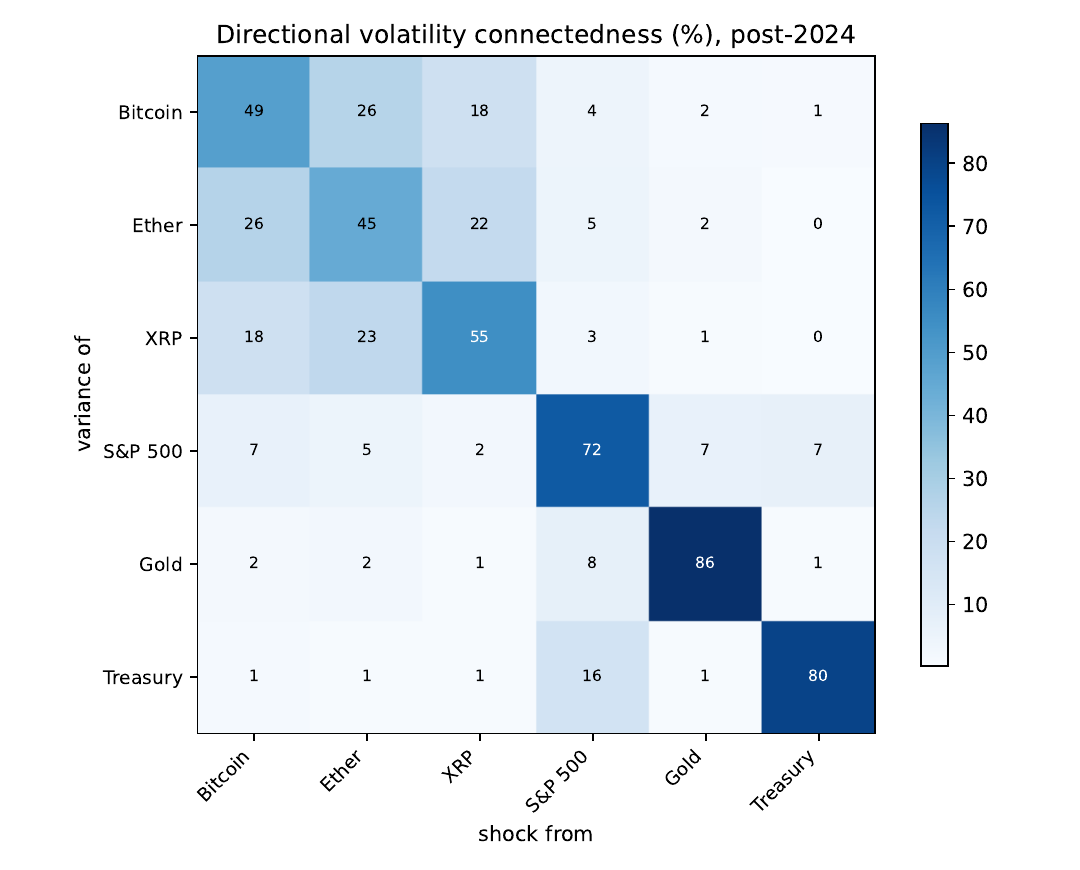}
\caption{Directional volatility connectedness (\%), post-2024. Row $i$, column $j$ gives the share of the
forecast-error variance of asset $i$ attributable to shocks in asset $j$, from the generalised variance
decomposition of \citet{dy2012}. Note the near block-diagonal structure and the small off-block entries.}
\label{fig:heat}
\end{figure}

We next apply the standard connectedness framework of \citet{dy2012} to the log realised variances of all six assets over the post-2024 sample. Figure~\ref{fig:heat} reports the resulting generalised variance decomposition.

The structure is approximately block-diagonal. The three cryptocurrencies form a tightly connected group, with 26\% of
Bitcoin's forecast-error variance attributable to Ether and 18\% to XRP. The three traditional assets are
similarly self-contained, each retaining between 72\% and 86\% of its own variance, the largest cross-link being
the 16\% of Treasury variance attributable to the S\&P~500. Transmission \emph{between} the blocks is small:
individual cryptocurrencies account for between 1\% and 7\% of the forecast-error variance of each traditional
asset. A rolling 250-day connectedness index shows no break at the spot-fund approvals, moving instead with
episodes of market stress, peaking near 43\% during the March~2020 pandemic shock and falling to about 27\% in
2022.

These results indicate limited average daily connectedness
between cryptocurrencies and traditional assets, but do not
rule out a relationship concentrated on particular days.
The standard daily connectedness framework does not explicitly
distinguish weekend cryptocurrency activity from weekday
activity or isolate its association with Monday variance.
A calendar-specific relationship may therefore be less
apparent in the aggregate decomposition. The weekend-channel analysis examines this possibility directly by relating $W_t$ to Monday log variance while controlling for the HAR terms and $F_t$. Its results complement Figure~\ref{fig:heat} by assessing a predictive relationship that the aggregate connectedness measure is not specifically designed to isolate.

\section{Conclusion}\label{sec:conc}

\subsection{Summary}
This paper documents a predictive relationship between weekend
cryptocurrency variance and Monday variance in the S\&P~500,
gold and a US Treasury bond fund. Weekend cryptocurrency
variance provides additional information beyond each asset's
own variance history and Friday cryptocurrency variance.
The relationship is weak before 2024 and significant across
all nine crypto-asset pairs in the post-2024 sample.
An endogenous sup-Wald test places the estimated break
in October~2023, preceding the January~2024 spot-Bitcoin
ETF approvals. This timing is compatible with anticipation
of approval, but does not establish its role in the change.

The weekend factor remains significant when the Friday
factor is excluded, whereas Friday cryptocurrency variance
alone has no significant predictive association. The
decomposition results support a predominantly common
cryptocurrency-market relationship, with no significant
additional predictive information detected in the
coin-specific residuals.

The estimated magnitude is substantial: a one-standard-deviation
increase in $W_t$ is associated with approximately
$27$-$54\%$ higher fitted Monday variance across the three
traditional assets, holding the HAR controls fixed.
Adding $W_t$ to the HAR benchmark reduces out-of-sample
QLIKE loss by approximately $11$-$22\%$, although the
statistical evidence for improved forecast accuracy
varies across markets. The findings generally remain
significant after controlling for Friday's VIX, allowing
for within-date dependence across pairs and removing
influential observations. The holiday-week results limit interpretation of the relationship as a generic reopening effect. ETF flow interactions also provide no support for a stronger
relationship on higher-flow Mondays. The evidence
therefore supports a Monday predictive relationship,
while its underlying mechanism remains unresolved.

\subsection{Interpretation}

Two aspects of the findings deserve emphasis. First, the aggregate connectedness results in Section~\ref{sec:conn} and the weekend-specific
regressions capture different features of cross-market relationships. Individual cryptocurrencies account for relatively small shares of traditional assets' forecast-error variance in the daily connectedness analysis, while the weekend factor provides additional predictive information about Monday variance. These findings are compatible because the daily connectedness framework does not explicitly isolate weekend cryptocurrency activity and its relationship with Monday outcomes. The comparison highlights the value of examining calendar-specific predictive relationships alongside aggregate connectedness.

Second, the emergence of the relationship around 2023-24 does not establish an ETF-related explanation. The estimated break precedes the January~2024
spot-Bitcoin ETF approvals, and the flow interactions do not support a stronger weekend relationship on Mondays with greater measured ETF flow activity. However, these findings do not exclude broader ETF-related or institutional channels. Overlapping ownership, dealer exposures and common risk-management practices remain possible explanations, but are not
directly tested here. The evidence supports a predictive relationship; its underlying mechanism remains unresolved.

\subsection{Limitations and further work}
Several limitations qualify these conclusions. Daily variance is estimated from OHLC prices rather than intraday returns. These range-based proxies measure
within-session price variation but exclude close-to-open price
changes and do not reveal when variation occurs within the session.
Intraday data would permit more detailed measurement and timing
analysis, subject to appropriate treatment of market microstructure
noise. The post-2024 estimation samples are short, containing 115–116 Mondays depending on the asset, so the emergence result rests on a limited number of observations and the endogenous break is only borderline significant at the 5\% level. The flow proxy is coarse and endogenous, so the
flow null should be read as an absence of evidence rather than strong evidence of absence. The fixed-income leg is
a 7-10 year bond fund rather than the on-the-run ten-year note. No cryptocurrency provides a clean never-listed
control by the end of the sample, since spot XRP funds launched in November~2025. The holiday-week results suggest a weaker weekend relationship when reopening is delayed until Tuesday, limiting a generic reopening interpretation. However, with only eleven holiday weeks, the evidence distinguishing a Monday-specific relationship from a generic reopening effect remains limited.

Four extensions follow naturally from these limitations. Intraday realised measures would sharpen estimation and allow the Monday effect
to be traced through the trading session rather than observed only daily. A longer post-event window will permit a
sharper test of the emergence date and reveal whether the channel persists, strengthens, or is gradually
arbitraged away as participants learn of it. Finer institutional data, such as dealer positions or fund holdings
rather than aggregate net flows, would allow the mechanism to be tested more precisely. Finally, the
identification strategy generalises: any pair of markets with differing closure schedules offers the same predictive setting based on differences in trading calendars, so the approach could be applied across time zones, or to assets whose closure lengths differ, to test
whether the length of the closed window governs the size of the effect.

\newpage
\appendix

\section{Robustness to the choice of range estimator}
\addcontentsline{toc}{section}{Appendix A. Robustness to the choice of range estimator}
\setcounter{table}{0}\renewcommand{\thetable}{A\arabic{table}}
\label{appendix}

The main analysis measures daily variance with the \citet{garmanklass1980} estimator. This appendix repeats
the principal tests using the two alternative range-based estimators introduced in Section~\ref{sec:rvmeasure}, that is,  the
\citet{parkinson1980} estimator, which uses only the high-low range, and the \citet{rogerssatchell1991}
estimator, which allows for a non-zero drift. For each estimator, every variance series is rebuilt from the
same daily prices with the same cleaning steps: the cryptocurrency weekend and Friday factors $W_t$ and $F_t$,
and the traditional asset's Monday log variance and HAR terms. The regression and forecasting exercises follow the corresponding main-text designs (Section~\ref{sec:results}), with all variance-based variables reconstructed using the chosen estimator. The Garman-Klass results are included as a benchmark.

\subsection*{How similar are the estimators?}

Table~\ref{tab:A1} reports correlations between the cleaned daily
log-variance series. Parkinson is closely correlated with
Garman-Klass for all six assets (0.96-0.98).
Rogers-Satchell is also highly correlated with Garman-Klass
for five assets (0.94-0.97), but less strongly for gold (0.75).
Before cleaning, Rogers-Satchell produces 367 zero estimates
for gold, compared with 131 under Garman-Klass and Parkinson.
Rogers-Satchell can equal zero despite a positive high-low range
when the opening and closing prices coincide with opposite
daily extremes. Zero estimates are treated as missing under
the cleaning procedure described in Section~\ref{sec:rvmeasure}.
The gold results therefore depend more heavily on this procedure,
particularly under Rogers-Satchell.

\begin{table}[H]\centering
\caption{Similarity of the alternative estimators, November 2017 to June 2026. Correlations are between cleaned daily log-variance series; the last two columns count days on which the estimator equals zero before cleaning.}
\label{tab:A1}
\begin{tabular}{lcccc}
\toprule
 & \multicolumn{2}{c}{Correlation with Garman-Klass} & \multicolumn{2}{c}{Days with zero estimate}\\
\cmidrule(lr){2-3}\cmidrule(lr){4-5}
Asset & Parkinson & Rogers-Satchell & \shortstack{Garman-Klass,\\Parkinson} & Rogers-Satchell\\
\midrule
Bitcoin & 0.97 & 0.97 & 0 & 5\\
Ether & 0.97 & 0.96 & 0 & 4\\
XRP & 0.98 & 0.97 & 0 & 2\\
S\&P 500 & 0.96 & 0.94 & 0 & 0\\
Gold & 0.96 & 0.75 & 131 & 367\\
Treasury (IEF) & 0.96 & 0.96 & 0 & 4\\
\bottomrule
\end{tabular}
\end{table}

\subsection*{The emergence of the relationship.}

Table~\ref{tab:A2} repeats the sub-period analysis of Table~\ref{tab:emerge}. All three estimators give a mean weekend
coefficient close to zero before 2020 and a clearly positive coefficient after January 2024, when eight or nine
of the nine pairs are positive and significant. The estimators differ in the intermediate sub-periods.
Parkinson identifies four positive and significant pairs in
2022-2023, whereas Rogers-Satchell identifies two in
2020-2021 and one in 2022-2023.

\begin{table}[H]\centering
\caption{Emergence of the weekend channel by estimator: mean of the nine pairwise weekend coefficients and the
number that are positive and significant at the 5\% level.}
\label{tab:A2}
\small
\begin{tabular}{lcccccc}
\toprule
 & \multicolumn{2}{c}{Garman-Klass} & \multicolumn{2}{c}{Parkinson} & \multicolumn{2}{c}{Rogers-Satchell}\\
\cmidrule(lr){2-3}\cmidrule(lr){4-5}\cmidrule(lr){6-7}
Sub-period & Mean $\hat\gamma$ & Sig. & Mean $\hat\gamma$ & Sig. & Mean $\hat\gamma$ & Sig.\\
\midrule
Pre-COVID (2017-11 to 2020-02) & $-0.03$ & 0/9 & $-0.04$ & 0/9 & $-0.02$ & 0/9\\
COVID-Recovery (2020-02 to 2021-12) & $0.03$ & 0/9 & $0.07$ & 0/9 & $0.02$ & 2/9\\
LUNA/FTX era (2022-01 to 2023-12) & $0.14$ & 2/9 & $0.19$ & 4/9 & $0.09$ & 1/9\\
Post-ETF (2024-01 to 2026-06) & $0.35$ & 9/9 & $0.29$ & 9/9 & $0.31$ & 8/9\\
\bottomrule
\end{tabular}
\end{table}

\subsection*{Weekend versus Friday, magnitude and forecasting value.}
Table~\ref{tab:A3} repeats Tables~\ref{tab:coll} and~\ref{tab:mag} for the post-2024 sample. Under every estimator and for every asset, the weekend coefficient
is positive and significant in the weekend-only specification
($p\leq0.021$), whereas the Friday-only coefficient is not
statistically significant ($p>0.30$). In the joint specification,
the one-sided test rejects $H_0:\gamma\leq\delta$ in favour of
$H_1:\gamma>\delta$ ($p\leq0.008$).

A one-standard-deviation increase in $W_t$ is associated with 27-54\% higher
fitted Monday variance under Garman-Klass, 27-46\% under Parkinson and 26-37\% under Rogers-Satchell.
Table~\ref{tab:A4} shows that the pooled panel coefficient remains between 0.31 and 0.38 with $t \ge 5.0$, and
that adding $W_t$ to HAR lowers out-of-sample QLIKE for all three assets under all three estimators, by 9-25\%.
As in the main text, the Diebold-Mariano evidence varies across markets: the one-sided $p$-values range from
0.024 to 0.094, and which asset reaches the 5\% level depends on the estimator.

\begin{table}[H]\centering
\caption{Weekend versus Friday and economic magnitude by estimator, post-2024. Weekend-only and Friday-only
specifications as in Table~\ref{tab:coll}; the contrast is the one-sided test of $\gamma>\delta$ in the joint model; the
one-standard-deviation effect and $\Delta R^2$ are computed as in Table~\ref{tab:mag}. Newey-West standard errors, five lags.}
\label{tab:A3}
\small
\begin{tabular}{llccccccc}
\toprule
 & & \multicolumn{2}{c}{Weekend-only} & \multicolumn{2}{c}{Friday-only} & Contrast & 1-SD & $\Delta R^2$\\
\cmidrule(lr){3-4}\cmidrule(lr){5-6}
Estimator & Asset & $\hat\gamma$ & $p$ & $\hat\delta$ & $p$ & $p$ & effect & (pp)\\
\midrule
\multirow{3}{*}{Garman-Klass} & S\&P 500 & 0.256 & 0.005 & $-0.015$ & 0.869 & 0.001 & $+28\%$ & 3.9\\
 & Gold & 0.447 & 0.001 & $-0.017$ & 0.883 & 0.001 & $+54\%$ & 6.2\\
 & Treasury (IEF) & 0.246 & 0.001 & $0.034$ & 0.521 & 0.001 & $+27\%$ & 7.0\\
\addlinespace
\multirow{3}{*}{Parkinson} & S\&P 500 & 0.238 & 0.009 & $0.007$ & 0.944 & 0.006 & $+27\%$ & 3.5\\
 & Gold & 0.377 & 0.008 & $0.035$ & 0.725 & 0.008 & $+46\%$ & 5.3\\
 & Treasury (IEF) & 0.248 & 0.001 & $0.036$ & 0.561 & $<$0.001 & $+28\%$ & 6.9\\
\addlinespace
\multirow{3}{*}{Rogers-Satchell} & S\&P 500 & 0.278 & 0.010 & $0.025$ & 0.783 & 0.002 & $+31\%$ & 3.9\\
 & Gold & 0.330 & 0.021 & $-0.117$ & 0.309 & $<$0.001 & $+37\%$ & 3.2\\
 & Treasury (IEF) & 0.235 & 0.003 & $0.014$ & 0.792 & 0.002 & $+26\%$ & 5.6\\
\bottomrule
\end{tabular}
\end{table}

\begin{table}[H]\centering
\caption{Pooled estimates, break dates and out-of-sample
forecast performance by variance estimator.}
\label{tab:A4}
\small
\begin{tabular}{lccccccc}
\toprule
 & Panel & \multicolumn{2}{c}{Endogenous break} & \multicolumn{3}{c}{QLIKE reduction (DM $p$)}\\
\cmidrule(lr){3-4}\cmidrule(lr){5-7}
Estimator & $\hat\gamma$ ($t$) & Date & sup-$\mathcal{W}$ & S\&P 500 & Gold & Treasury\\
\midrule
Garman-Klass & 0.38 (5.6) & Oct 2023 & 8.8 & 22\% (0.066) & 11\% (0.050) & 17\% (0.042)\\
Parkinson & 0.36 (5.3) & Dec 2022 & 6.8 & 15\% (0.047) & 15\% (0.063) & 18\% (0.056)\\
Rogers-Satchell & 0.31 (5.0) & Apr 2024 & 10.1 & 25\% (0.094) & 9\% (0.024) & 12\% (0.060)\\
\bottomrule
\end{tabular}
\end{table}

\begin{minipage}{\textwidth}
\footnotesize
\textit{Notes:} Panel regressions standardise the outcome and
crypto predictors within each pair, control for the daily HAR
term, and cluster standard errors by date; parentheses contain
$t$-statistics. Break regressions include all HAR terms but
exclude Friday variance, with 15\% trimming. Forecast entries
show QLIKE reductions, with one-sided DM $p$-values in parentheses.
\end{minipage}

\subsection*{Timing of the change.}
The estimated break date is sensitive to the choice of estimator.
The maximum sup-Wald statistic occurs in October 2023 under Garman-Klass ($8.8$), December 2022 under Parkinson ($6.8$), and April 2024 under Rogers-Satchell ($10.1$). The alternative estimators therefore give different estimates
of the timing of the change. These differences do not establish
whether the underlying change was abrupt or gradual.

\subsection*{Implications.}
The positive post-2024 relationship is robust to the choice of
range estimator. Under both alternatives, the weekend factor
is positive and statistically significant for all three traditional
assets in the weekend-only specifications controlling for HAR
terms. Friday-only coefficients are not statistically significant,
and joint-model contrasts favour a larger weekend than Friday
coefficient. Estimated economic effects remain substantial.
Adding the weekend factor reduces out-of-sample QLIKE for every
asset under each estimator, although statistical evidence of
forecast improvement varies across assets and estimators.

The timing of the change is less robust. No pairwise coefficient
is positive and statistically significant in the pre-COVID
sub-period, compared with eight or nine in the post-ETF
sub-period. Estimated break dates differ across estimators.
The October 2023 estimate reported in Section~\ref{sec:emerge} should
therefore be interpreted cautiously rather than as a precisely
established onset. These results do not establish whether
spot-ETF approvals caused the stronger relationship.


\newpage
\begin{thebibliography}{99}
\small
\bibitem[Andersen and Bollerslev(1998)]{andersenbollerslev1998} Andersen, T.G., Bollerslev, T., 1998. Answering the skeptics: yes, standard volatility models do provide accurate forecasts. \textit{International Economic Review} 39, 885-905.
\bibitem[Andersen et al.(2001)]{abde2001} Andersen, T.G., Bollerslev, T., Diebold, F.X., Ebens, H., 2001. The distribution of realized stock return volatility. \textit{Journal of Financial Economics} 61, 43-76.
\bibitem[Andersen et al.(2003)]{abdl2003} Andersen, T.G., Bollerslev, T., Diebold, F.X., Labys, P., 2003. Modeling and forecasting realized volatility. \textit{Econometrica} 71, 579-625.
\bibitem[Andrews(1993)]{andrews1993} Andrews, D.W.K., 1993. Tests for parameter instability and structural change with unknown change point. \textit{Econometrica} 61, 821-856.
\bibitem[Antonakakis et al.(2020)]{acg2020} Antonakakis, N., Chatziantoniou, I., Gabauer, D., 2020. Refined measures of dynamic connectedness based on time-varying parameter vector autoregressions. \textit{Journal of Risk and Financial Management} 13, 84.
\bibitem[Babalos et al.(2025)]{babalos2025} Babalos, V., Bouri, E., Gupta, R., 2025. Does the introduction of US spot Bitcoin ETFs affect spot returns and volatility of major cryptocurrencies? \textit{Quarterly Review of Economics and Finance} 102, 102006.
\bibitem[Barndorff-Nielsen and Shephard(2002)]{bns2002} Barndorff-Nielsen, O.E., Shephard, N., 2002. Econometric analysis of realized volatility and its use in estimating stochastic volatility models. \textit{Journal of the Royal Statistical Society: Series B} 64, 253-280.
\bibitem[Barun\'ik and K\v{r}ehl\'ik(2018)]{bk2018} Barun\'ik, J., K\v{r}ehl\'ik, T., 2018. Measuring the frequency dynamics of financial connectedness and systemic risk. \textit{Journal of Financial Econometrics} 16, 271-296.
\bibitem[Baur and Dimpfl(2018)]{baurdimpfl2018} Baur, D.G., Dimpfl, T., 2018. Asymmetric volatility in cryptocurrencies. \textit{Economics Letters} 173, 148-151.
\bibitem[Baur and Lucey(2010)]{baurlucey2010} Baur, D.G., Lucey, B.M., 2010. Is gold a hedge or a safe haven? An analysis of stocks, bonds and gold. \textit{Financial Review} 45, 217-229.
\bibitem[Bollerslev(1986)]{bollerslev1986} Bollerslev, T., 1986. Generalized autoregressive conditional heteroskedasticity. \textit{Journal of Econometrics} 31, 307-327.
\bibitem[Bouri et al.(2017)]{bouri2017} Bouri, E., Moln\'ar, P., Azzi, G., Roubaud, D., Hagfors, L.I., 2017. On the hedge and safe haven properties of Bitcoin: is it really more than a diversifier? \textit{Finance Research Letters} 20, 192-198.
\bibitem[Breiman(2001)]{breiman2001} Breiman, L., 2001. Random forests. \textit{Machine Learning} 45, 5-32.
\bibitem[Bucci(2020)]{bucci2020} Bucci, A., 2020. Realized volatility forecasting with neural networks. \textit{Journal of Financial Econometrics} 18, 502-531.
\bibitem[Bukaita and Li(2026)]{bukaita2026} Bukaita, W., Li, X., 2026. Structural spillovers among Bitcoin, Ethereum, gold, and U.S. equities: evidence from the 2024 spot ETF institutionalization regime. \textit{Economies} 14(4), 143.
\bibitem[Caporale and Plastun(2019)]{caporale2019} Caporale, G.M., Plastun, A., 2019. The day of the week effect in the cryptocurrency market. \textit{Finance Research Letters} 31, 258-269.
\bibitem[Chen and Guestrin(2016)]{chenguestrin2016} Chen, T., Guestrin, C., 2016. XGBoost: a scalable tree boosting system. In: \textit{Proceedings of the 22nd ACM SIGKDD International Conference on Knowledge Discovery and Data Mining}, 785-794.
\bibitem[Cont(2001)]{cont2001} Cont, R., 2001. Empirical properties of asset returns: stylized facts and statistical issues. \textit{Quantitative Finance} 1, 223-236.
\bibitem[Corbet et al.(2019)]{corbet2019} Corbet, S., Lucey, B., Urquhart, A., Yarovaya, L., 2019. Cryptocurrencies as a financial asset: a systematic analysis. \textit{International Review of Financial Analysis} 62, 182-199.
\bibitem[Corsi(2009)]{corsi2009} Corsi, F., 2009. A simple approximate long-memory model of realized volatility. \textit{Journal of Financial Econometrics} 7, 174-196.
\bibitem[Cross(1973)]{cross1973} Cross, F., 1973. The behavior of stock prices on Fridays and Mondays. \textit{Financial Analysts Journal} 29(6), 67-69.
\bibitem[Diebold and Mariano(1995)]{dm1995} Diebold, F.X., Mariano, R.S., 1995. Comparing predictive accuracy. \textit{Journal of Business \& Economic Statistics} 13, 253-263.
\bibitem[Diebold and Yilmaz(2012)]{dy2012} Diebold, F.X., Yilmaz, K., 2012. Better to give than to receive: predictive directional measurement of volatility spillovers. \textit{International Journal of Forecasting} 28, 57-66.
\bibitem[Diebold and Yilmaz(2014)]{dy2014} Diebold, F.X., Yilmaz, K., 2014. On the network topology of variance decompositions. \textit{Journal of Econometrics} 182, 119-134.
\bibitem[Dudek et al.(2024)]{dudek2024} Dudek, G., Fiszeder, P., Kubus, P., Orzeszko, W., 2024. Forecasting cryptocurrencies volatility using statistical and machine learning methods: a comparative study. \textit{Applied Soft Computing} 151, 111132.
\bibitem[Dyhrberg(2016)]{dyhrberg2016} Dyhrberg, A.H., 2016. Bitcoin, gold and the dollar: a GARCH volatility analysis. \textit{Finance Research Letters} 16, 85-92.
\bibitem[Engle(1982)]{engle1982} Engle, R.F., 1982. Autoregressive conditional heteroscedasticity with estimates of the variance of United Kingdom inflation. \textit{Econometrica} 50, 987-1007.
\bibitem[Fama(1965)]{fama1965} Fama, E.F., 1965. The behavior of stock-market prices. \textit{Journal of Business} 38, 34-105.
\bibitem[French(1980)]{french1980} French, K.R., 1980. Stock returns and the weekend effect. \textit{Journal of Financial Economics} 8, 55-69.
\bibitem[French and Roll(1986)]{frenchroll1986} French, K.R., Roll, R., 1986. Stock return variances: the arrival of information and the reaction of traders. \textit{Journal of Financial Economics} 17, 5-26.
\bibitem[Friedman(2001)]{friedman2001} Friedman, J.H., 2001. Greedy function approximation: a gradient boosting machine. \textit{Annals of Statistics} 29, 1189-1232.
\bibitem[Garman and Klass(1980)]{garmanklass1980} Garman, M.B., Klass, M.J., 1980. On the estimation of security price volatilities from historical data. \textit{Journal of Business} 53, 67-78.
\bibitem[Glosten et al.(1993)]{gjr1993} Glosten, L.R., Jagannathan, R., Runkle, D.E., 1993. On the relation between the expected value and the volatility of the nominal excess return on stocks. \textit{Journal of Finance} 48, 1779-1801.
\bibitem[Gu et al.(2020)]{gukellyxiu2020} Gu, S., Kelly, B., Xiu, D., 2020. Empirical asset pricing via machine learning. \textit{Review of Financial Studies} 33, 2223-2273.
\bibitem[Hansen and Lunde(2005)]{hansenlunde2005} Hansen, P.R., Lunde, A., 2005. A forecast comparison of volatility models: does anything beat a GARCH(1,1)? \textit{Journal of Applied Econometrics} 20, 873-889.
\bibitem[Hochreiter and Schmidhuber(1997)]{hochreiter1997} Hochreiter, S., Schmidhuber, J., 1997. Long short-term memory. \textit{Neural Computation} 9, 1735-1780.
\bibitem[Hong et al.(2025)]{hong2025} Hong, Y., Feng, H., Wang, Y., Li, B., 2025. The impact of Bitcoin ETF approval on Bitcoin's hedging properties against traditional assets. Working paper, arXiv:2512.12815.
\bibitem[IMF(2022)]{imf2022} International Monetary Fund, 2022. Cryptic connections: spillovers between crypto and equity markets. \textit{Global Financial Stability Notes} 2022/01.
\bibitem[Iyer and Popescu(2023)]{iyer2023} Iyer, R., Popescu, A., 2023. New evidence on spillovers between crypto assets and financial markets. \textit{International Monetary Fund Working Paper}, 23/213.
\bibitem[Jahanshahloo et al.(2022)]{Jahanshahloo2022} Jahanshahloo, H., Corbet, S., Oxley, L., 2022. Seeking sigma: Time-of-the-day effects on the Bitcoin network. \textit{Finance Research Letters} 49, 103101.
\bibitem[Katsiampa(2017)]{katsiampa2017} Katsiampa, P., 2017. Volatility estimation for Bitcoin: a comparison of GARCH models. \textit{Economics Letters} 158, 3-6.
\bibitem[Mandelbrot(1963)]{mandelbrot1963} Mandelbrot, B., 1963. The variation of certain speculative prices. \textit{Journal of Business} 36, 394-419.
\bibitem[Mourey et al.(2025)]{mourey2025} Mourey, M., Shahrour, M.H., \c{S}oiman, F., 2025. A crypto-stock weekend effect: predicting Monday stock returns using weekend cryptocurrency returns. \textit{Finance Research Letters} 86, 108661.
\bibitem[Nelson(1991)]{nelson1991} Nelson, D.B., 1991. Conditional heteroskedasticity in asset returns: a new approach. \textit{Econometrica} 59, 347-370.
\bibitem[Newey and West(1987)]{nw1987} Newey, W.K., West, K.D., 1987. A simple, positive semi-definite, heteroskedasticity and autocorrelation consistent covariance matrix. \textit{Econometrica} 55, 703-708.
\bibitem[Parkinson(1980)]{parkinson1980} Parkinson, M., 1980. The extreme value method for estimating the variance of the rate of return. \textit{Journal of Business} 53, 61-65.
\bibitem[Patton(2011)]{patton2011} Patton, A.J., 2011. Volatility forecast comparison using imperfect volatility proxies. \textit{Journal of Econometrics} 160, 246-256.
\bibitem[Rahimikia and Poon(2020)]{rahimikia2020} Rahimikia, E., Poon, S.-H., 2020. Machine learning for realised volatility forecasting. Working paper, SSRN 3707796.
\bibitem[Rogers and Satchell(1991)]{rogerssatchell1991} Rogers, L.C.G., Satchell, S.E., 1991. Estimating variance from high, low and closing prices. \textit{Annals of Applied Probability} 1, 504-512.
\bibitem[SEC(2024)]{sec2024} U.S. Securities and Exchange Commission, 2024. Statement on the approval of spot Bitcoin exchange-traded products. 10 January 2024, Washington, DC.
\bibitem[Urquhart(2016)]{urquhart2016} Urquhart, A., 2016. The inefficiency of Bitcoin. \textit{Economics Letters} 148, 80-82.
\end{thebibliography}
\end{document}